\documentclass[12pt]{article}

\usepackage{cite}
\usepackage{epsfig}
\usepackage{amsmath}
\usepackage{amssymb}
\usepackage{cancel}
\usepackage{pstricks}
\usepackage{afterpage}

\newcommand{\gs}{g_{\rm s}}
\newcommand{\as}{\alpha_{\rm s}}
\def\mathswitch#1{\relax\ifmmode#1\else$#1$\fi}
\newcommand{\msbar}{\mathswitch{\overline{\text{MS}}}}
\newcommand{\tev}{\,\, \mathrm{TeV}}

\newcommand{\mycaption}[1]{\caption{\sl #1}}

\makeatletter
\def\section{\@startsection {section}{1}{\z@}{+3.0ex plus +1ex minus
  +.2ex}{2.3ex plus .2ex}{\large\bf\boldmath}}
\def\subsection{\@startsection{subsection}{2}{\z@}{+2.5ex plus +1ex
minus +.2ex}{1.5ex plus .2ex}{\normalsize\bf\boldmath}}
\def\subsubsection{\@startsection{subsubsection}{3}{\z@}{+3.25ex plus
 +1ex minus +.2ex}{1.5ex plus .2ex}{\normalsize\it}}

\graphicspath{{.}{plots/}}

\begin{document}
\thispagestyle{empty}

\def\thefootnote{\fnsymbol{footnote}}

\begin{flushright}
\end{flushright}

\vspace{1cm}

\begin{center}

{\Large {\bf QCD corrections to massive color-octet\\[.5ex] vector boson pair
production}}
\\[3.5em]
{\large
Ayres~Freitas and Daniel Wiegand
}

\vspace*{1cm}

{\sl
Pittsburgh Particle-physics Astro-physics \& Cosmology Center
(PITT-PACC),\\ Department of Physics \& Astronomy, University of Pittsburgh,
Pittsburgh, PA 15260, USA
}

\end{center}

\vspace*{2.5cm}

\begin{abstract}

This paper describes the calculation of the next-to-leading order (NLO) QCD
corrections to massive color-octet vector boson pair production at hadron
colliders.
As a concrete framework, a two-site coloron model with an internal parity is
chosen, which can be regarded as an effective low-energy approximation of
Kaluza-Klein gluon physics in universal extra dimensions. The renormalization procedure
involves several subtleties, which are discussed in detail. The impact of the
NLO corrections is relatively modest, amounting to a reduction of 11--14\% in
the total cross-section, but they significantly
reduce the scale dependence of the LO result.

\end{abstract}

\setcounter{page}{0}
\setcounter{footnote}{0}

\newpage


\section{Introduction}

\noindent
Massive color-octet vector bosons appear in a number of
beyond-the-Standard-Model (BSM) theories, such as universal extra dimensions
(UED) \cite{ued,uedreview}, topcolor models \cite{topcolor}, 
coloron models \cite{coloron}, and moose models \cite{moose}. They may be
copiously produced at the Large Hadron Collider (LHC), leading to distinct
signatures \cite{colpheno} that are actively searched for \cite{colatlas,colcms}. Most of these
analyses consider single resonance production of the massive octet vectors, with
decays into dijet or top-pair final states.

On the other hand, single production of massive color-octet vector bosons is forbidden
or suppressed in UED models with Kaluza-Klein (KK) parity or in moose models with
a $\mathbb{Z}_2$ exchange symmetry, so that pair production becomes the leading
production process. The phenomenology of these particles at the LHC has been
studied extensively, see for example Ref.~\cite{uedreview,uedpheno}. However,
these analyses were based on tree-level predictions for the relevant production
cross-sections, which are subject to large uncertainties from QCD radiative
corrections.

QCD corrections have been computed for a number of pair production processes
of colored BSM particles, including (but not limited to) squark and gluino
production in the Minimal Supersymmetric Standard Model (MSSM)
\cite{sqgl,sqgl2,sqglres}, leptoquark pair production \cite{lq}, production of massive
vector quarks \cite{vq}, and pair production of scalar color-octet bosons
\cite{sgluon}. The corrections were generically found to be sizeable and
important to reduce the large dependence of tree-level results on the
renormalization scale. Thus, for a robust prediction of the production of
colored BSM particles at hadron colliders, the inclusion of next-to-leading
order (NLO) QCD corrections is mandatory.

QCD corrections to production of single vector octets have been studied in
Refs.~\cite{colqcd,colqcd2}. In this paper, we consider pair production of
massive color-octet vector bosons at hadron colliders at NLO precision. For
concreteness, the calculation is based on a two-site coloron model with exchange
symmetry. This model can be regarded as a low-energy effective theory of minimal
UED with one extra dimension (mUED), which includes only the first KK level as
dynamic degrees of freedom. In contrast to new colored scalars or fermions, the
analysis of colored vector bosons involves several subtleties concerning the
gauge fixing and the renormalization procedure. In particular, there is an
inherent ambiguity in the definition of the coupling renormalization. This is a
reflection of the fact that the two-site model is manifestly non-renormalizable
and thus depends on assumptions about the ultra-violet (UV) completion. This
issue will be discussed in some detail in the following, before presenting the
technical aspects of the calculation and the numerical results.

The paper is organized as follows: In the next section, the two-site coloron
model is introduced, including a detailed description of the role of the
exchange symmetry, which is reminiscent of KK-parity in mUED. Section~\ref{sc:calc}
discusses the calculation of the 
NLO corrections to coloron pair production. Special emphasis is placed on the
renormalization procedure and the treatment of infra-red (IR) divergencies
through phase-space slicing. In section~\ref{sc:res}, numerical results for the
total cross-section and the rapidity distribution are shown, before concluding
in section~\ref{sc:concl}. For the reader's convenience, the Feynman rules of
the two-site coloron model are provided in the appendix.


\section{The two-site symmetric coloron model}
\label{sc:model}

The two-site coloron model is based on an extension of the strong gauge group to
the product group $\text{SU(3)}_1 \times \text{SU(3)}_2$, which is broken down
to $\text{SU(3)}_{\rm C}$ by a non-linear sigma model. In addition, invariance
under the $\mathbb{Z}_2$ transformation $\cal P$ is imposed, which interchanges 
the two SU(3) groups:
\begin{align}
{\cal P}: \quad \text{SU(3)}_1 \leftrightarrow \text{SU(3)}_2. 
\end{align}
This exchange
symmetry mimics the KK parity of UED. The Lagrangian of the
model can be divided into three parts,
\begin{align}
{\cal L} = {\cal L}_{\rm gauge} + {\cal L}_{\rm ferm} + {\cal L}_{\rm gf}.
\end{align}
The gauge part is given by
\begin{align}
{\cal L}_{\rm gauge} = 
 -\frac{1}{4} G_{1\mu\nu}G_1^{\mu\nu}
 -\frac{1}{4} G_{2\mu\nu}G_2^{\mu\nu}
 + \frac{f^2}{4} \, \text{tr}\{ D_\mu\Sigma D^\mu \Sigma^\dagger \}.
\label{eq:Lgauge}
\end{align}
Here $G_{i\mu\nu}$ are the field strength tensors of SU(3)$_i$ ($i=1,2$), with gauge
couplings $g_1=g_2 \equiv g$. $\Sigma$ denotes the non-linear sigma field
\begin{align}
\Sigma = \exp(2i\pi^A T^A/f),
\end{align}
where $A=1,...,8$ is implicitly summed over, $T^A$ are the SU(3)
generators, $f$ is a constant of mass dimension, and $\pi^A$ are the Goldstone
fields of the broken SU(3). Its covariant derivative is given by
\begin{align}
D_\mu\Sigma = \partial_\mu\Sigma -ig\, G^A_{1\mu} T^A \Sigma
+ ig\, \Sigma G^A_{2\mu} T^A.
\end{align}
Under $\text{SU(3)}_1 \times \text{SU(3)}_2$, the $\Sigma$ field
transforms as a bi-fundamental,
\begin{align}
\Sigma \to U_1 \, \Sigma \, U_2^\dagger.
\end{align}
The $\Sigma$ field is responsible for the breaking of
$\text{SU(3)}_1 \times \text{SU(3)}_2$ to the vectorial subgroup
$\text{SU(3)}_{\rm C}$. The gauge mass eigenstates in the broken phase are
\begin{align}
G^A_\mu &= \tfrac{1}{\sqrt{2}} (G^A_{1\mu} + G^A_{2\mu}), &
C^A_\mu &= \tfrac{1}{\sqrt{2}} (G^A_{1\mu} - G^A_{2\mu}).
\end{align}
Here $G^A_\mu$ is the (massless) gluon field of $\text{SU(3)}_{\rm C}$ with coupling strength
$\gs = g/\sqrt{2}$, whereas $C^A_\mu$ is the massive coloron field with
mass $M = \gs f$, which ``eats'' the Goldstone fields $\pi^A$.

Eq.~\eqref{eq:Lgauge} has the same form as 
for the coloron model in Ref.~\cite{colqcd} with the additional
constraint that the two gauge groups have equal coupling strength. The latter
requirement is a consequence of the $\cal P$ parity, which was not considered in
Ref.~\cite{colqcd}. Under this parity
\begin{align}
{\cal P}: \quad G_{1\mu}^A &\leftrightarrow G_{2\mu}^A, &
G_\mu^A &\to G_\mu^A, &
C_\mu^A &\to -C_\mu^A, &
\Sigma &\to \Sigma^\dagger .
\end{align}
Since $C^A_\mu$ is odd under $\cal P$, the massive colorons can only be produced
in pairs.

The fermion part of the Lagrangian reads
\begin{align}
{\cal L}_{\rm ferm} = 
{}&  \bar{q}_1 i \cancel{D}_1 q_1 
+ \bar{q}_2 i \cancel{D}_2 q_2 
+ \bar{q}' i \cancel{D}_{\rm V} q'
- Y \bigl [ \bar{q}_1 \xi q' - \bar{q}_2 \xi^\dagger q' + \text{h.c.} \bigr ]
 \notag \\
&+ \bar{u}_1 i \cancel{D}_1 u_1 
+ \bar{u}_2 i \cancel{D}_2 u_2 
+ \bar{u}' i \cancel{D}_{\rm V} u'
- Y \bigl [ \bar{u}_1 \xi u' - \bar{u}_2 \xi^\dagger u' + \text{h.c.} \bigr ]
 \label{eq:Lferm} \\
&+ \bar{d}_1 i \cancel{D}_1 d_1 
+ \bar{d}_2 i \cancel{D}_2 d_2 
+ \bar{d}' i \cancel{D}_{\rm V} d'
- Y \bigl [ \bar{d}_1 \xi d' - \bar{d}_2 \xi^\dagger d' + \text{h.c.} \bigr ].
 \notag
\end{align}
Here $\psi_1$, $\psi_2$ and $\psi'$ are quark fields in the fundamental representation of
SU(3)$_1$, SU(3)$_2$ and SU(3)$_{\rm C}$, respectively ($\psi=q,u,d$). The
$\psi=q$
fields are chiral doublets under the weak SU(2)$_{\rm W}$ group, whereas
$\psi=u,d$ are singlets. The relevant quantum numbers and chirality of the quark fields is
summarized in Tab.~\ref{tab:quarks}. 
\begin{table}[t]
\centering
\begin{tabular}{lccccc}
\hline
Field & Chirality & SU(2)$_{\rm W}$ & SU(3)$_1$ & SU(3)$_2$ & SU(3)$_{\rm C}$ \\
\hline
$q_1$ & L & \bf 2 & \bf 3 & \bf 1 & -- \\
$q_2$ & L & \bf 2 & \bf 1 & \bf 3 & -- \\
$q'$ & R & \bf 2 & -- & -- & \bf 3 \\
\hline
$u_1$ & R & \bf 1 & \bf 3 & \bf 1 & -- \\
$u_2$ & R & \bf 1 & \bf 1 & \bf 3 & -- \\
$u'$ & L & \bf 1 & -- & -- & \bf 3 \\
\hline
$d_1$ & R & \bf 1 & \bf 3 & \bf 1 & -- \\
$d_2$ & R & \bf 1 & \bf 1 & \bf 3 & -- \\
$d'$ & L & \bf 1 & -- & -- & \bf 3 \\
\hline
\end{tabular}
\mycaption{Quantum numbers and chirality of the quark fields in the two-site
symmetric coloron model.
\label{tab:quarks}}
\end{table}
Their covariant derivatives read
\begin{align}
D_{1\mu} \psi_1 &= \partial_\mu \psi_1 - i g G^A_{1\mu} T^A \psi_1 + ..., \notag \\
D_{1\mu} \psi_2 &= \partial_\mu \psi_2 - i g G^A_{2\mu} T^A \psi_2 + ..., \notag 
 & [\psi = q,u,d] \\
D_{\rm V\mu} \psi' &= \partial_\mu \psi' - \tfrac{i g}{\sqrt{2}} 
 (G^A_{1\mu}+G^A_{2\mu}) T^A \psi' + ...,
\end{align}
where the dots indicate electroweak interactions, which are ignored in this
work.
Furthermore, $\xi$ is the ``square root'' sigma field according to the CCWZ
construction \cite{ccwz},
\begin{align}
\xi = \exp(i\pi^A T^A/f).
\end{align}
Under $\text{SU(3)}_1 \times \text{SU(3)}_2$, these fields transform as
\begin{align}
\psi_1^{} &\to U_1^{} \psi_1^{}, &
\psi_2^{} &\to U_2^{} \psi_2^{}, &
\psi' &\to U_{\rm V}^{} \psi', &
\xi \to U_1^{}\xi U_{\rm V}^\dagger = U_{\rm V}^{} \xi U_2^\dagger,
\end{align}
where $U_{\rm V}$ is the transformation matrix of the fundamental representation
of the vectorial subgroup SU(3)$_{\rm C}$. The effect of $\cal P$ parity on the
fermion fields is
\begin{align}
{\cal P}: \quad \psi_1 &\leftrightarrow \psi_2, & 
\psi' &\to -\psi', & \xi &\to \xi^\dagger.
\end{align}
The introduction of the $\psi'$ fields is necessary to be able to write down
invariant Yukawa terms (with coupling strength $Y$) in eq.~\eqref{eq:Lferm}.

The physical quark mass eigenstates are 
\begin{align}
\psi &= \tfrac{1}{\sqrt{2}}(\psi_1 + \psi_2), && [\psi = q,u,d] \notag \\
\Psi &= \tfrac{1}{\sqrt{2}}(\psi_1 - \psi_2)P_L + \psi'P_R, 
 && [\Psi = Q,U,D]
\end{align}
where $P_{L,R} = \frac{1}{2}(1\pm \gamma_5)$. Here the $\psi$ fields are massless
chiral $\cal P$-even SM-like quark fields, whereas the $\Psi$ fields are $\cal P$-odd
fermion fields with a vector-like mass $M_\Psi = \sqrt{2} Y$. In general, the Yukawa
coupling $Y$ is a free parameter, but for the sake of analogy to UED we impose
\begin{align}
Y = M/\sqrt{2}, \quad \text{i.e.} \quad M_\Psi = M.
\end{align}
For the top quark, the SM Higgs Yukawa coupling cannot be ignored. It leads to
mixing between the $U_3$ and first component of the $Q_3$ fields, where the
subscript indicates the generation index, see e.g.\
App.~H of Ref.~\cite{uedcomphep}. The mass matrix reads
\begin{align}
\begin{pmatrix}
\overline{Q}_3 & \overline{U}_3
\end{pmatrix}
\begin{pmatrix}
M & m_t \\ m_t & -M
\end{pmatrix}
\begin{pmatrix}
Q_3 \\ U_3
\end{pmatrix},
\end{align}
leading to two degenerate mass eigenstates $T$ and $T'$ given by
\begin{align}
\begin{pmatrix}
Q_3 \\ U_3
\end{pmatrix}
= 
\begin{pmatrix}
\cos \theta_T & \gamma_5 \sin \theta_T \\
\sin\theta_T & - \gamma_5 \cos \theta_T
\end{pmatrix}
\begin{pmatrix}
T \\ T'
\end{pmatrix}
\label{eq:tmix}
\end{align}
with mass and mixing angle
\begin{align}
M_T &= \sqrt{M^2+m_t^2},  & \tan 2\theta_T &= \frac{m_t}{M}.
\label{eq:tpars}
\end{align}
The final component of the model is the gauge-fixing and ghost term. 
For a covariant gauge it can be defined in the following $\cal P$-symmetric form,
\begin{align}
{\cal L}_{\rm gf} &= 
 -\tfrac{1}{2}({\cal F}_1^A)^2 - \tfrac{1}{2}({\cal F}_2^A)^2 + \sum_{i,j=1}^2
 \bar{u}^A_i \frac{\delta {\cal F}^A_i}{\delta \alpha^B_j} u^B_j,
 \label{eq:Lgf}
\intertext{where}
{\cal F}_1^A &= \frac{1}{\sqrt{\xi}} G^A_{1\mu} 
 + \sqrt{\xi} \frac{g}{2}f \,\pi^A, \notag \\
{\cal F}_2^A &= \frac{1}{\sqrt{\xi}} G^A_{2\mu} 
 - \sqrt{\xi} \frac{g}{2}f \,\pi^A,
\end{align}
and $\delta\alpha^A_i$ is the parameter of an infinitesimal SU(3)$_i$ gauge
transformation.
For the calculation presented in the following sections, the Feynman gauge
$\xi=1$ has been employed. In this gauge, the unphysical Goldstone fields
$\pi^A$ receive a mass $M = \gs f = gf/\sqrt{2}$ from eq.~\eqref{eq:Lgf}. The
ghost fields mix to form a $\cal P$-even massless gluon ghost $u_g =
\frac{1}{\sqrt{2}}(u_1+u_2)$ and a $\cal P$-odd coloron ghost $u_C  =
\frac{1}{\sqrt{2}}(u_1+u_2)$ with mass $M$. Thus one obtains
\begin{align}
{\cal L}_{\rm gf} = &{} -\frac{1}{2} \Bigl [
(\partial^\mu G_\mu^A)^2 + (\partial^\mu C_\mu^A)^2 \Bigr ]
- \frac{M^2}{2} (\pi^A)^2 - M\, \partial^\mu C_\mu^A \,\pi^A \notag \\
&- \bar{u}^A_g \partial^2 u^A_g - \bar{u}^A_C (\partial^2+M^2) u^A_C
+ \gs f_{ABC} \,\bar{u}_g^A \partial^\mu (u_g^B G_\mu^C + u_C^B C_\mu^C) \notag \\
&+ \gs f_{ABC} \,\bar{u}_C^A \partial^\mu (u_g^B C_\mu^C + u_C^B G_\mu^C)
+ \gs M f_{ABC} (\bar{u}^A_g u^B_C - \bar{u}^A_C u^B_g) \pi^C.
\end{align}
In summary, the two-site symmetric coloron model defined in this way contains
several states with mass $M$ in addition to the SM particle content. Besides the
coloron vector-boson,  heavy vector-like quarks are required to enforce the
$\cal P$-parity as an exact symmetry.

This model can be viewed as a low-energy approximation of the 5-dimensional
minimal UED
model (mUED) with compactification radius $R=M^{-1}$, where only the zero modes and first KK
excitations are kept as dynamical degrees of freedom. Note, however, that the
coloron model is not identical to a simple truncation of mUED
at the $N_{\rm KK}=1$ level, since such a truncated UED model would violate
gauge invariance \cite{uedgv}, whereas the model presented here respects the
full gauge symmetry, albeit non-linearly.
In fact, the Feynman rules for the two-site coloron model and
the first KK excitation in mUED are mostly identical, but there are a few
differences, which are mentioned in appendix~\ref{sc:feynr}.

In a more general sense, the two-site symmetric coloron model can be regarded as
a low-energy description of any model with massive color-octet vector bosons
that are odd under some (approximate) parity.


\section{NLO corrections to the pair production process}
\label{sc:calc}

\begin{figure}[t]
\centering
\includegraphics[width=5in]{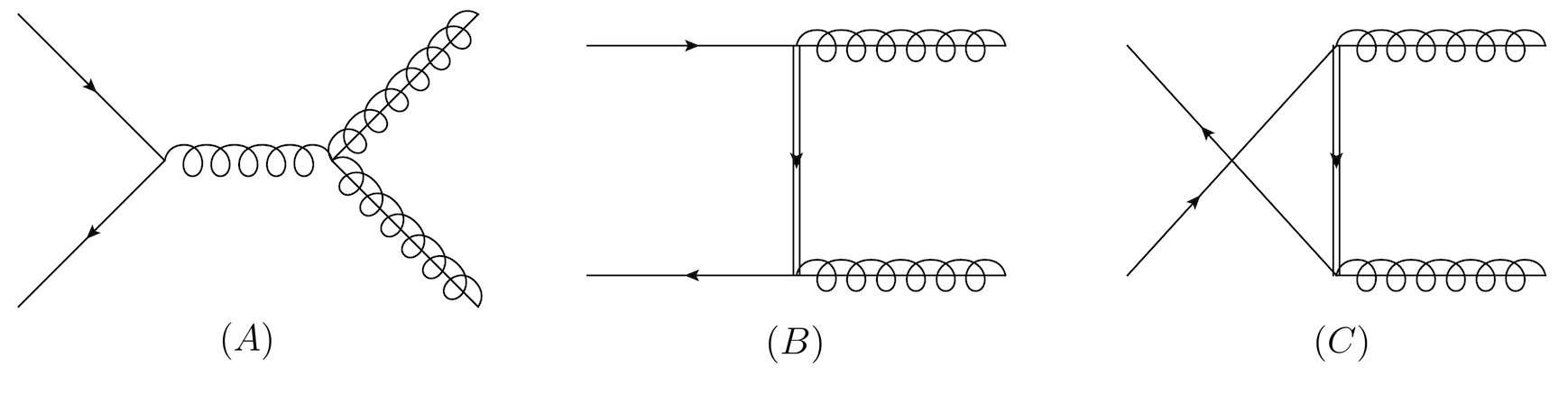}\\
\includegraphics[width=5in]{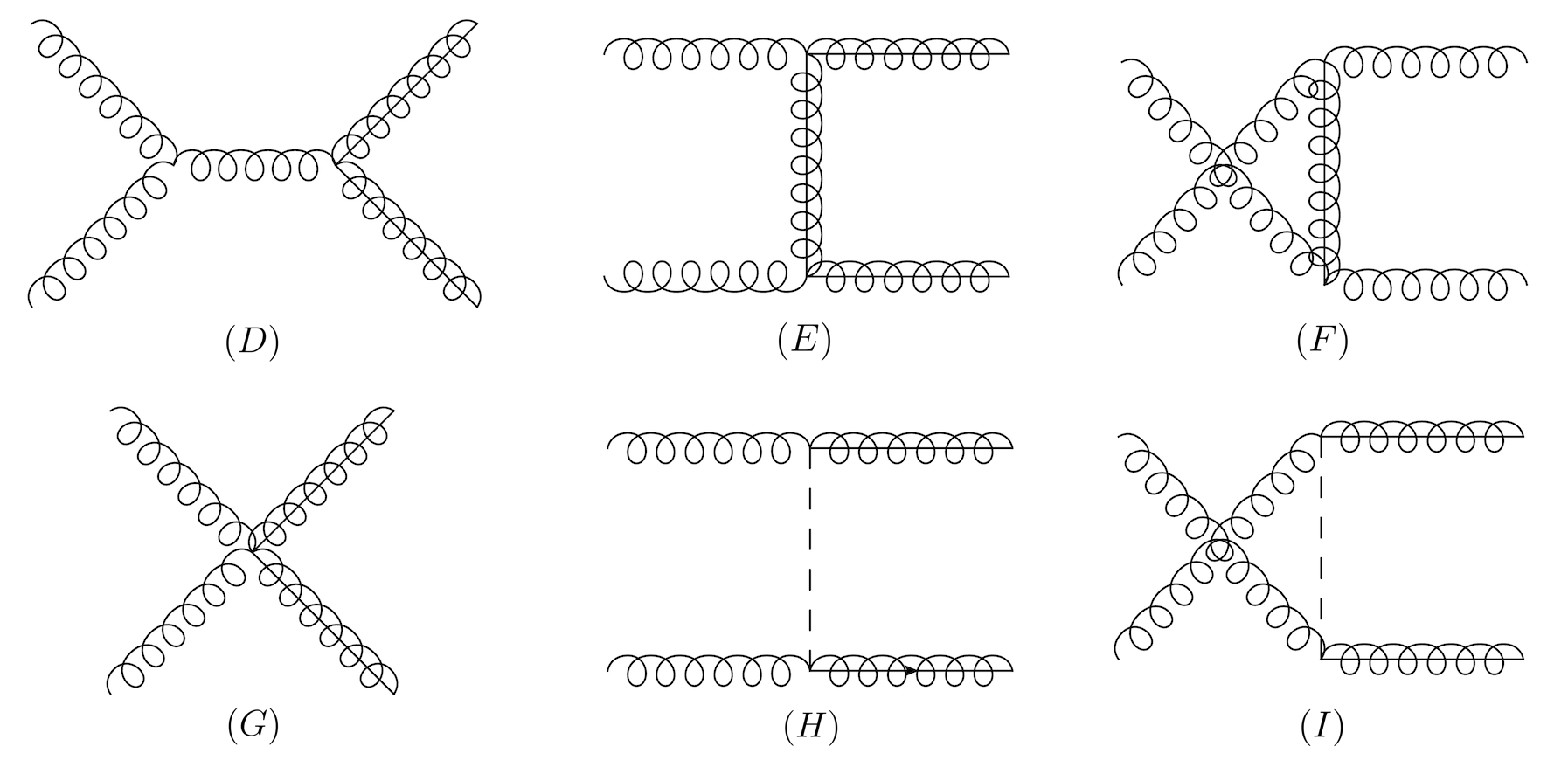}
\mycaption{Born-level diagrams contributing to massive color-octet vector-boson
pair production. Here the spring--solid lines indicate massive color-octet
vector-bosons, while the double lines indicate massive $\cal P$-odd quarks, and
the dashed line indicates a $\cal P$-odd Goldstone scalar.
\label{fig:diag1}}
\end{figure}

\noindent
Massive colorons can be pair produced at hadron colliders, such as the LHC. The
tree-level process $pp \to CC$ can be divided into two partonic sub-channels,
$q\bar{q} \to CC$ and $gg \to CC$, with the relevant diagrams shown in
Fig.~\ref{fig:diag1}. Note that at leading order this process is identical to
of KK gluon pair production in mUED.

\begin{figure}[tb]
\centering
\includegraphics[width=4in]{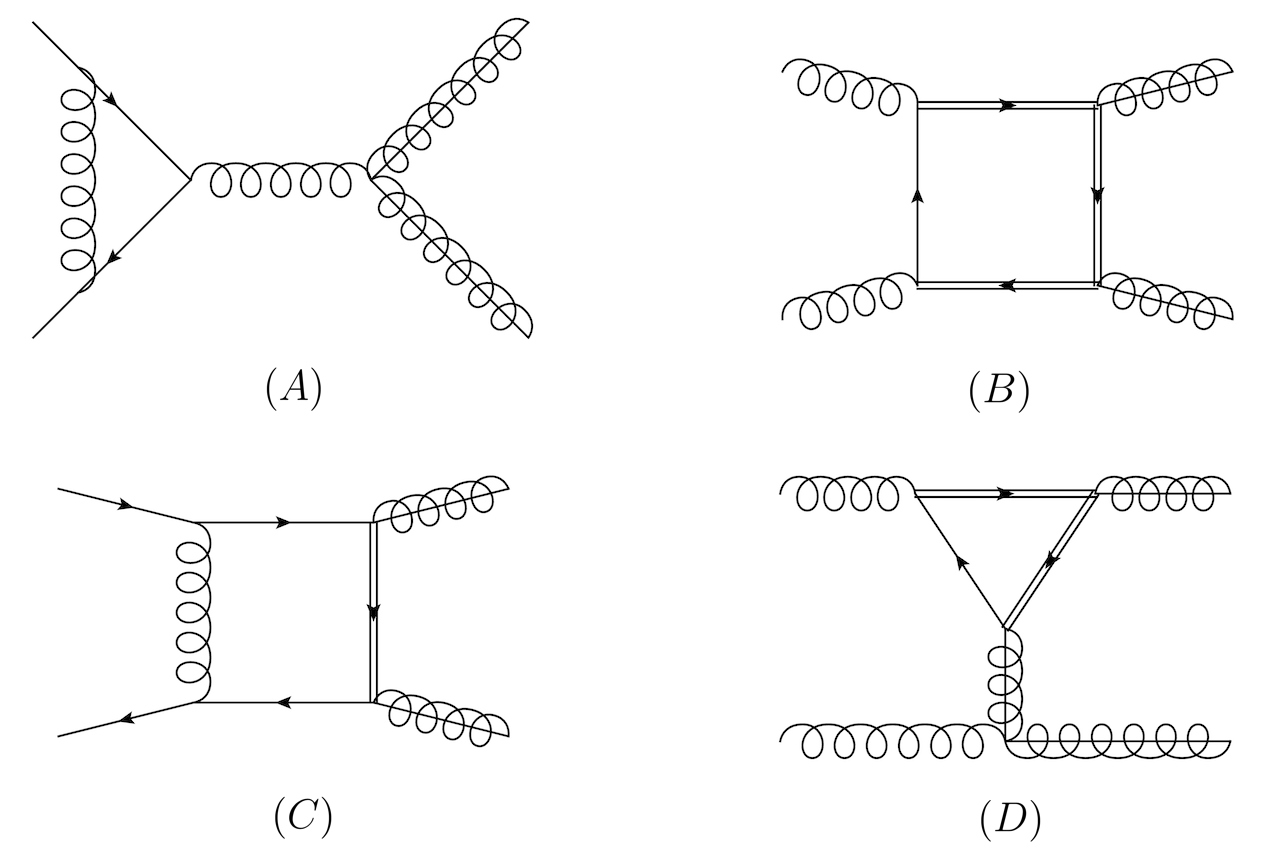}
\mycaption{Sample one-loop diagrams contributing to coloron pair production. See
Fig.~\ref{fig:diag1} for the definition of the different propagator line types.
\label{fig:diag2}}
\end{figure}

\begin{figure}[t]
\vspace{1em}
\centering
\includegraphics[width=5in]{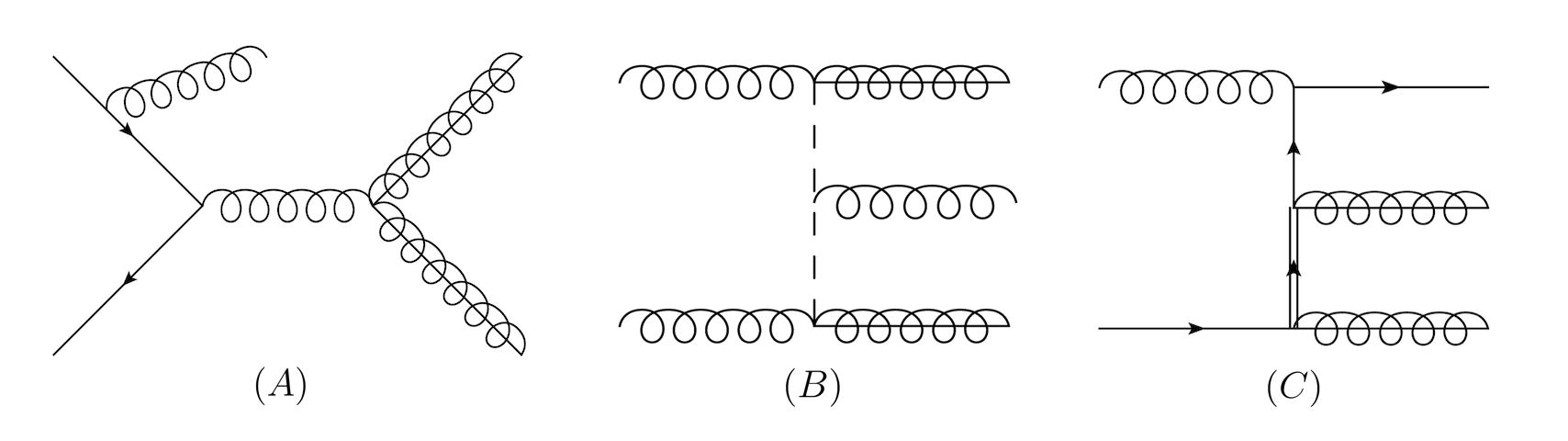}
\vspace{-1ex}
\mycaption{Sample real radiation diagrams contributing to coloron pair production. See
Fig.~\ref{fig:diag1} for the definition of the different propagator line types.
\label{fig:diag3}}
\end{figure}

At NLO, one needs to consider one-loop corrections to the subprocesses $q\bar{q}
\to CC$ and $gg \to CC$, as well as real emission of an extra gluon at
tree-level, $q\bar{q} \to CCg$ and $gg \to CCg$. A few sample diagrams are shown
in Figs.~\ref{fig:diag2} and \ref{fig:diag3}. Both the loop contributions and
real emission contributions are separately IR divergent, but the divergencies
cancel in the combined result. Additionally, the quark-gluon induced
subprocesses $qg\to CCq$ and $\bar{q}g \to CC\bar{q}$ appear for the first time
at NLO.


At NLO, the predictions for coloron pair production become sensitive to
assumptions about the UV completion. The renormalization procedure employed here
takes a bottom-up approach, assuming that the
running couplings are defined at the mass scale $M$ of the colorons\footnote{If
instead the couplings are defined at a high scale $\Lambda \gg M$, this may lead
to additional moderately-sized contributions to the NLO result. This will be
explored in future work. However, experience from other BSM calculations
indicates that the numerically dominant part of the NLO QCD is generated by SM
gluon exchange contributions and thus does not depend on the details of the UV
completion.}.
In the next subsection, the renormalization scheme is discussed in more detail.


\subsection{Renormalization}
\label{sc:renorm}

\noindent
In this work, the renormalization is performed by using the on-shell scheme for
the wave-function and mass renormalization of the physical states and \msbar\
renormalization for the strong coupling constant. However, due to the fact that
the two-site coloron model is fundamentally a non-renormalizable theory, there
are several subtleties that need to be addressed. These will be discussed in
this section, together with a brief summary of the remaining aspects of the
renormalization.

\medskip\noindent
For the external states the wave-function renormalization constants
\begin{align}
\delta Z^\psi_{\rm L} = 
\delta Z^\psi_{\rm R} \quad [\psi=q,u,d], \qquad
\delta Z^g, \qquad
\delta Z^C
\end{align}
are introduced for the left- and right-handed (massless) SM quarks, the gluons,
and the massive colorons, respectively. As usual, their values are determines
through the residues of the renormalized propagators, leading to
\begin{align}
\delta Z^\psi_{\rm L,R} &= -{\Re\rm e}\{ \Sigma^\psi_{\rm L,R}(0)\}, &
\delta Z^g &= -{\Re\rm e}\bigl\{ \tfrac{\partial}{\partial(p^2)}\Sigma^g(0)
\bigr\}, &
\delta Z^C &= -{\Re\rm e}\bigl\{ \tfrac{\partial}{\partial(p^2)}\Sigma^C(M^2)
\bigr\},
\end{align}
where $\Sigma^\psi_{\rm L,R}(p^2)$, $\Sigma^g(p^2)$ and $\Sigma^C(p^2)$ are the
left/right-handed quark self-energies, transverse gluon self-energy and
transverse coloron self-energy, respectively.

The masses of the colorons and massive quarks are renormalized according to the
on-shell prescriptions
\begin{align}
\delta M_C^2 &= {\Re\rm e}\{\Sigma^C(M^2)\}, &
\delta M_\Psi &= \frac{M}{2} {\Re\rm e}\bigl\{
 \Sigma^\psi_{\rm L}(M^2) + \Sigma^\psi_{\rm R}(M^2) + 
 2\Sigma^\psi_{\rm S}(M^2) \bigr\}.
\end{align}
The mass parameter in the gauge-fixing term gets renormalized in the same way as
the coloron mass.
Note that, while we assume that the colorons and massive quarks have the same
mass $M$ at tree-level, as in mUED, they are technically independent parameters
in the coloron model and thus receive different mass counterterms. In mUED, in
fact, the degeneracy of the KK masses is also broken at the one-loop level due
to boundary terms \cite{uedmass}.

Following the analogy to mUED, therefore, we assume that the mass
difference between the coloron mass, $M_C$, and the vector-like quark mass,
$M_\Psi$, is small: $|M_C-M_\Psi|/M \sim {\cal O}(\as)$. Within the
contributions to ${\cal O}(\as)$ we thus set $M_C=M_\Psi=M$ but allow the masses to
deviate by a small
numerical amount in the tree-level contribution, consistent with this power
counting.

\medskip\noindent
The strong coupling constant is renormalized in the 5-flavor \msbar\ scheme. In
this scheme, only the gluons and five light quarks are included in the scale
evolution of the $\as$, whereas the scale dependence of the top quark, coloron
and heavy vector quark loops is accounted for through explicit logarithms in the
finite part of the counterterm. See e.g.\ Ref.~\cite{sqgl} for an application of
this scheme in the context of supersymmetry. For the $g\psi\bar{\psi}$, $ggg$,
$g\Psi\bar{\Psi}$ and $gCC$ gauge coupling, this leads to
\begin{align}
&\gs^{\rm bare} \to \gs(\mu)\,\bigl (1+ \delta Z_g \bigr ) \\
&\begin{aligned}[b]
\delta Z_g =  \frac{\as(\mu)}{4\pi} \biggl [ \!
 &-\frac{\beta_0}{2} \biggl ( \frac{1}{\epsilon} - \gamma_{\rm E} + \log(4\pi)
  \biggr ) \\
 &-\frac{1}{3}\log\frac{m_t^2}{\mu^2}
 +\biggl (\frac{21}{4} - \frac{2}{3}n_q \biggr ) \log \frac{M^2}{\mu^2}
 -\frac{2}{3} \log\frac{M_T^2}{\mu^2}
\biggr ],
\end{aligned} \\[1ex]
&\beta_0 = \beta_0^{\rm L} + \beta_0^{\rm H}
 = \Bigl ( 11 - \frac{2}{3} n_q \Bigr ) + \Bigl (\frac{21}{2}-
 \frac{4n_q+6}{3} \Bigr ),
\end{align}
where $n_q=5$, and $\mu$ is the renormalization scale, which is taken equal to the
regularization scale for simplicity. Furthermore, $\epsilon = (4-d)/2$, where
$d$ is the number of dimensions in dimensional regularization.

On the other hand, for the $C\psi\Psi$ and $gC\pi$ couplings (where $\pi$ is a
Goldstone boson), one needs different coupling counterterms. This is not
entirely surprising, since these couplings are not SU(3)$_{\rm C}$ gauge
interactions, but are instead related to the larger non-linear $\text{SU(3)}_1
\times \text{SU(3)}_2$ symmetry. 

To determine the $\mu$-dependence of these couplings, one may assume that all
gluon and coloron coupling have the same value at $\mu = M$, and the $C\psi\Psi$
and $gC\pi$ couplings do not effectively run for $\mu < M$. Thus one finds
\begin{align}
&C\psi\Psi: &&\delta Z'_g = \frac{\as(\mu)}{4\pi} \biggl [ 
 -\biggl (10 - \frac{2}{3}n_q \biggr )
  \biggl ( \frac{1}{\epsilon} - \gamma_{\rm E} + \log(4\pi) \biggr )
  +\frac{\beta_0^{\rm L}}{2} \log \frac{M^2}{\mu^2} \biggr ], \\[1ex]
&gC\pi: &&\delta Z''_g = \frac{\as(\mu)}{4\pi} \biggl [ 
 -\biggl (\frac{21}{4} - \frac{n_q}{2} \biggr )
 \biggl ( \frac{1}{\epsilon} - \gamma_{\rm E} + \log(4\pi) \biggr )
  +\frac{\beta_0^{\rm L}}{2} \log \frac{M^2}{\mu^2} \biggr ].
\end{align}

\medskip\noindent
In addition, one needs a counterterm for the vacuum expectation value of the
sigma field, $\Sigma$. This counterterm, denoted by the symbol $\delta t$,
appears in the renormalization of the Goldstone self-energy:
\begin{align}
\includegraphics[width=.9in]{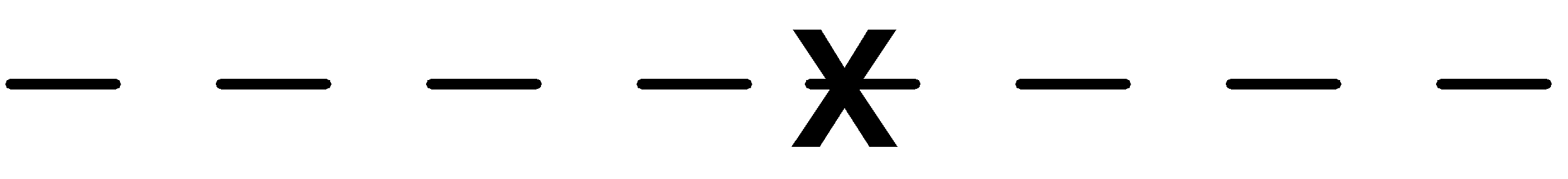}\; = i\delta_{AB} \frac{\delta t}{M}.
\end{align}
In a Higgs-like theory, this counterterm is usually determined from the
requirement that the renormalized tadpole terms of the Higgs field should
vanish. For the coloron model, however, the symmetry breaking mechanism is left
unspecified, and the radial degrees of freedom  of the sigma field (which
correspond to the Higgs scalars in a weakly coupled symmetry breaking sector)
are assumed to be integrated out. Therefore the tadpole condition cannot be used
here.

On the other hand, as explained for example in Ref.~\cite{ilisie}, the
counterterm for the vacuum expectation also appears in the Goldstone
self-energy $\Sigma^\pi(p^2)$. 
Thus one can impose the renormalization condition
\begin{align}
\delta t = -M \, \Sigma^\pi(0),
\end{align}
which in a Higgs-like theory is completely equivalent to the tadpole condition.


\subsection{Cancellation of IR divergencies}
\label{sc:ir}

\noindent
The real radiation contributions contain divergencies from soft and collinear
gluon emission, which cancel against the corresponding singularities in the
virtual loop contributions. To carry out this cancellation explicitly, the
phase-space slicing method with two cutoffs is employed here
\cite{Harris:2001sx}. According to this scheme, the phase space integration of
the $2\to 3$ real radiation contribution is split into three categories,
\begin{align}
\sigma_{2 \to 3} = \frac{1}{2s} \int d\Gamma_3 \, |{\cal M}_3|^2
 = \frac{1}{2s} \biggl [
 \int_{\rm S} d\Gamma_3 \, |{\cal M}_3|^2 +
 \int_{\rm C} d\Gamma_3 \, |{\cal M}_3|^2 +
 \int_{\rm H} d\Gamma_3 \, |{\cal M}_3|^2 \biggr ].
\end{align}
Here $d\Gamma_3$ is the three-particle phase-space measure, and ${\cal M}_3$ is
the $2\to 3$ matrix element. On the right-hand side,
``S'' indicates the soft region, where the gluon energy is restricted to
\begin{align}
0 \leq E_g \leq \delta_{\rm s} \frac{\sqrt{\hat s}}{2},
\end{align}
where $\hat{s}$ is the partonic center-of-mass energy.
For sufficiently small values of $\delta_{\rm s}$, the soft contribution
factorizes into the born matrix element and an eikonal factor,
\begin{align}
\int_{\rm S} d\Gamma_3 \, |{\cal M}_3|^2
&= \int d\Gamma_2 \, |{\cal M}_2|^2 \times
\frac{\as}{2\pi} \, \frac{\Gamma(1-\epsilon)}{\Gamma(1-2\epsilon)}\Bigl (
 \frac{4\pi\mu_{\rm R}^2}{s}\Bigr )^\epsilon
\sum_{i,j}\int d\Gamma_g\, \frac{-p_i\cdot p_j}{(p_i \cdot p_g)(p_j \cdot p_g)}
\,.
\label{eq:softfac}
\end{align}
Here $d\Gamma_2$ and ${\cal M}_2$ are the two-particle phase-space measure and
Born matrix element, respectively, while $d\Gamma_g$ is the single-particle
phase-space measure for the gluon momentum, and the sum $\sum_{i,j}$ runs over
all external legs. The eikonal factor can be
integrated analytically (see e.g.\ Refs.~\cite{Harris:2001sx,Beenakker:1988bq}).

The label ``C'' denotes the hard collinear region, defined by
\begin{align}
&\delta_{\rm s} \frac{\sqrt{\hat s}}{2} < E_g, &
&1-\cos\theta_{gi} \leq \delta_{\rm c}\frac{\sqrt{\hat s}}{E_g},
\end{align}
where $\theta_{gi}$ is the angle between the final-state gluon and the incoming
parton $i$ ($i=1,2$). For small $\delta_{\rm c}$, the phase space measure and
matrix element factorize into the born contribution and the divergent
Altarelli-Parisi splitting kernels. In dimensional regularization one thus
obtains
\begin{align}
\int_{\rm C} d\Gamma_3 \, |{\cal M}_3|^2
&= \int d\Gamma_2 \, |{\cal M}_2|^2 \times
\frac{\as}{2\pi} \, \frac{\Gamma(1-\epsilon)}{\Gamma(1-2\epsilon)} \Bigl (
 \frac{4\pi\mu_{\rm R}^2}{s}\Bigr )^\epsilon \Bigl (\frac{A_1^{\rm c}}{\epsilon} +
 A_0^{\rm c} \Bigr ), \label{eq:collfac}
\end{align}
where $A_1^{\rm c}$ and $A_0^{\rm c}$ are
known numerical constants (see e.g.\ Refs.~\cite{Harris:2001sx}). The collinear
divergencies in the splitting functions can be absorbed into the renormalization
of the parton distribution functions (PDFs) of the incoming partons. The form of
eq.~\eqref{eq:collfac} presumes that the $\msbar$ scheme, with the
renormalization scale $\mu_{\rm R}$, is used for  this purpose.

The soft and collinear contributions are combined with the virtual corrections
to arrive at
\begin{align}
d\sigma &= \sum_{i,j} \int dx_1 dx_2 \, \Bigl\{
 \bigl [ f_i(x_1,\mu_{\rm F}) f_j(x_2,\mu_{\rm F}) + 
 (1 \leftrightarrow 2) \bigr ] \bigl [d\hat{\sigma}^{(0)}_{ij}(\hat{s})
 + d\hat{\sigma}^{(1)}_{ij}(\hat{s};\delta_{\rm s})\bigr ] \notag \\
&\qquad +
 \bigl [ \tilde{f}_i(x_1,\mu_{\rm F}) f_j(x_2,\mu_{\rm F}) + 
  \tilde{f}_j(x_1,\mu_{\rm F}) f_i(x_2,\mu_{\rm F}) + (1 \leftrightarrow 2)
  \bigr ] d\hat{\sigma}^{(0)}_{ij}(\hat{s}) \Bigr\}
\label{eq:2to2}
\intertext{with}
\tilde{f}_i(x,\mu_{\rm F}) &=
 \sum_k \int_{x}^{1-\delta_{\rm s}} \frac{dz}{z} \;
  f_k\Bigl(\frac{x}{z},\mu_{\rm F}\Bigr) \, \frac{\as}{2\pi}
  \biggl [ P_{ik}(z) \,\ln \biggl ( \frac{\hat{s}}{\mu_{\rm F}^2}\,
  \frac{1-z}{z} \, \delta_{\rm c} \biggr ) - P'_{ik}(z) \Biggr ].
\end{align}
Here $f_i(x,\mu_{\rm F})$ is the proton PDF for the parton $i$ with the
factorization scale $\mu_{\rm F}$; $d\hat{\sigma}^{(0)}_{ij}$ is the differential
partonic Born cross-section for the incoming partons $i$ and $j$; 
$d\hat{\sigma}^{(0)}_{ij}$ is the one-loop corrected partonic cross-section
including the soft radiation terms; $P_{ik}(z)$ and $P'_{ik}(z)$ are the finite
and ${\cal O}(\epsilon)$ pieces of the unregulated splitting kernels (see e.g.\
Refs.~\cite{Harris:2001sx}), and $\hat{s} = x_1x_2s$.
Note that the form of eq.~\eqref{eq:2to2} changes slightly for the quark-gluon
induced subprocesses, which do not receive Born contributions.

The remaining hard radiation region, labeled ``H'', is constrained by the
conditions $\delta_{\rm s} \frac{\sqrt{\hat s}}{2} < E_g$ and
$1-\cos\theta_{gi} > \delta_{\rm c}\frac{\sqrt{\hat s}}{E_g}$. It is finite and can be
computed with numerical Monte-Carlo integration methods. Both the hard
contribution and the result in eq.~\eqref{eq:2to2} separately depend on the choices for
$\delta_{\rm s}$ and $\delta_{\rm c}$. However, as long as the cutoff parameters
are kept sufficiently small, this dependence drops out in the combined total result.


\subsection{Notes on the technical implementation}
\label{sc:impl}

\noindent
The calculation has been performed using several publicly available computing
tools, but additional components were specifically implemented by the authors.
The Feynman rules of the coloron model (see Appendix \ref{sc:feynr}) have been
incorporated into {\sc FeynArts 3} \cite{feynarts}, which was used for
generating the relevant diagrams and amplitudes. The color, Dirac and Lorentz
algebra was performed with {\sc FeynCalc} \cite{feyncalc}.

To simplify the treatment of tensor loop integrals, the one-loop amplitude was
contracted with the Born amplitude and the sum over the spins of external
particles carried out before any tensor reduction. As a result, most tensor
structures in the numerator of the loop integrand can be canceled against
propagator denominators. For the remaining tensor integrals, Passarino-Veltman
reduction has been used \cite{pv}. One thus arrives at a final result in terms
of standard one-loop basis functions. The IR-finite basis integrals have been
evaluated numerically using {\sc LoopTools 2} \cite{looptools}, whereas the
IR-divergent basis integrals were taken from Ref.~\cite{ez}.

For the $q\bar{q}$ channel, two fully independent calculations have been carried
out. One is based on dimensional regularization for the UV singularities and gluon
and quark mass regulators for the soft and collinear divergencies, respectively.
The other calculation has employed dimensional regularization for all types of
singularities. Perfect agreement between the two results at the level of
differential cross-sections was obtained. For the $gg$ channel, the use of a mass
regulator is not suitable. Nevertheless, we have performed many independent
checks of partial contributions to the final result.

The numerical integration over the final-state phase space and initial-state
PDFs is implemented in the form of a Monte-Carlo generator in Fortran.
This implementation is based on Ref.~\cite{Buckley:2014ana} and produces weighted parton-level
events.


\section{Numerical results}
\label{sc:res}

\noindent
In the following, we present phenomenological results for coloron pair
production at the LHC with $\sqrt{s}=14\tev$. Throughout this section, the
CTEQ6.1M PDF set \cite{cteq} have been used, as incorporated in the LHAPDF
framework \cite{lhapdf}.

\begin{figure}[t]
\centering
\includegraphics[height=6.2cm]{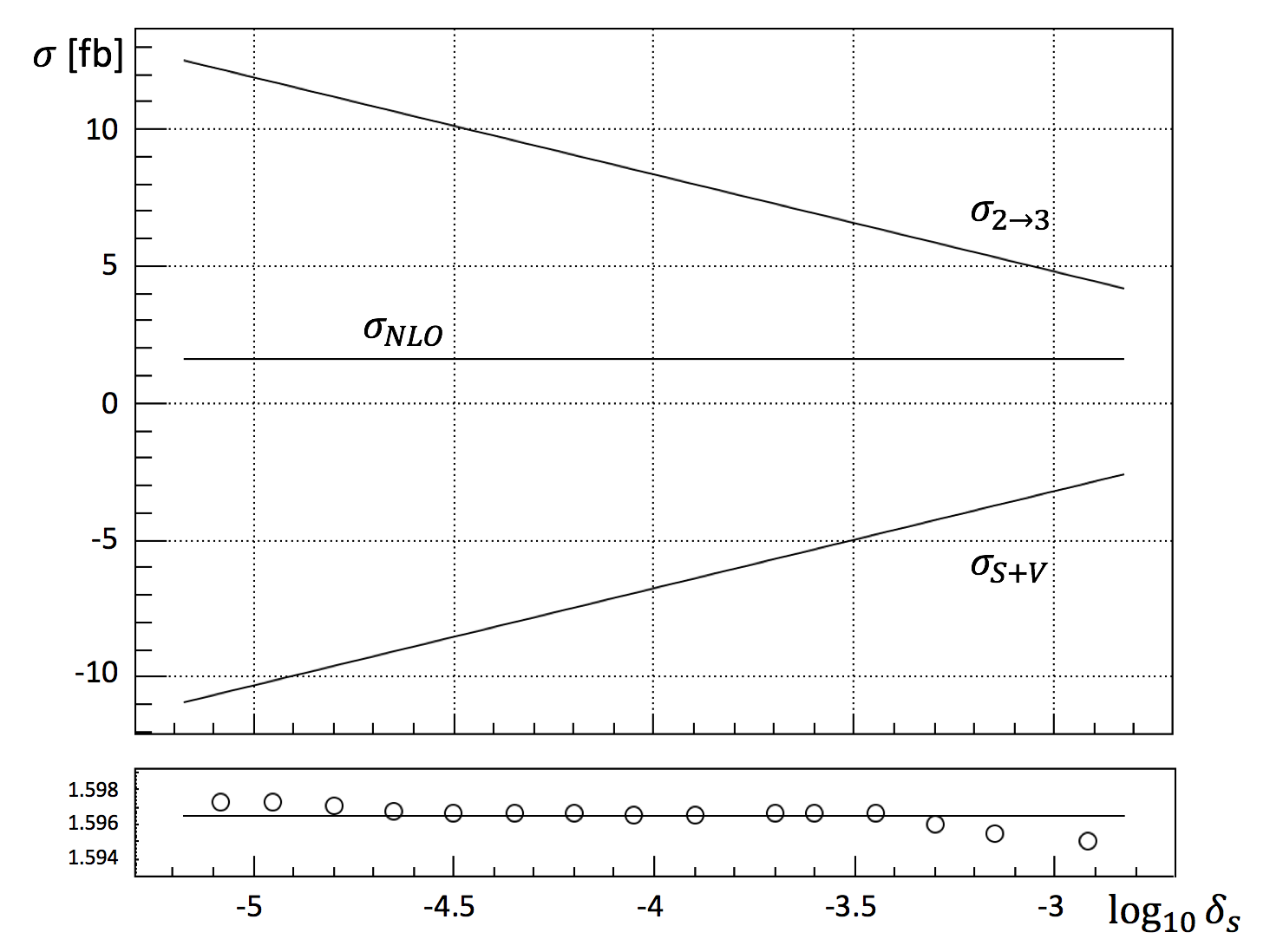}\hfill
\includegraphics[height=6.2cm]{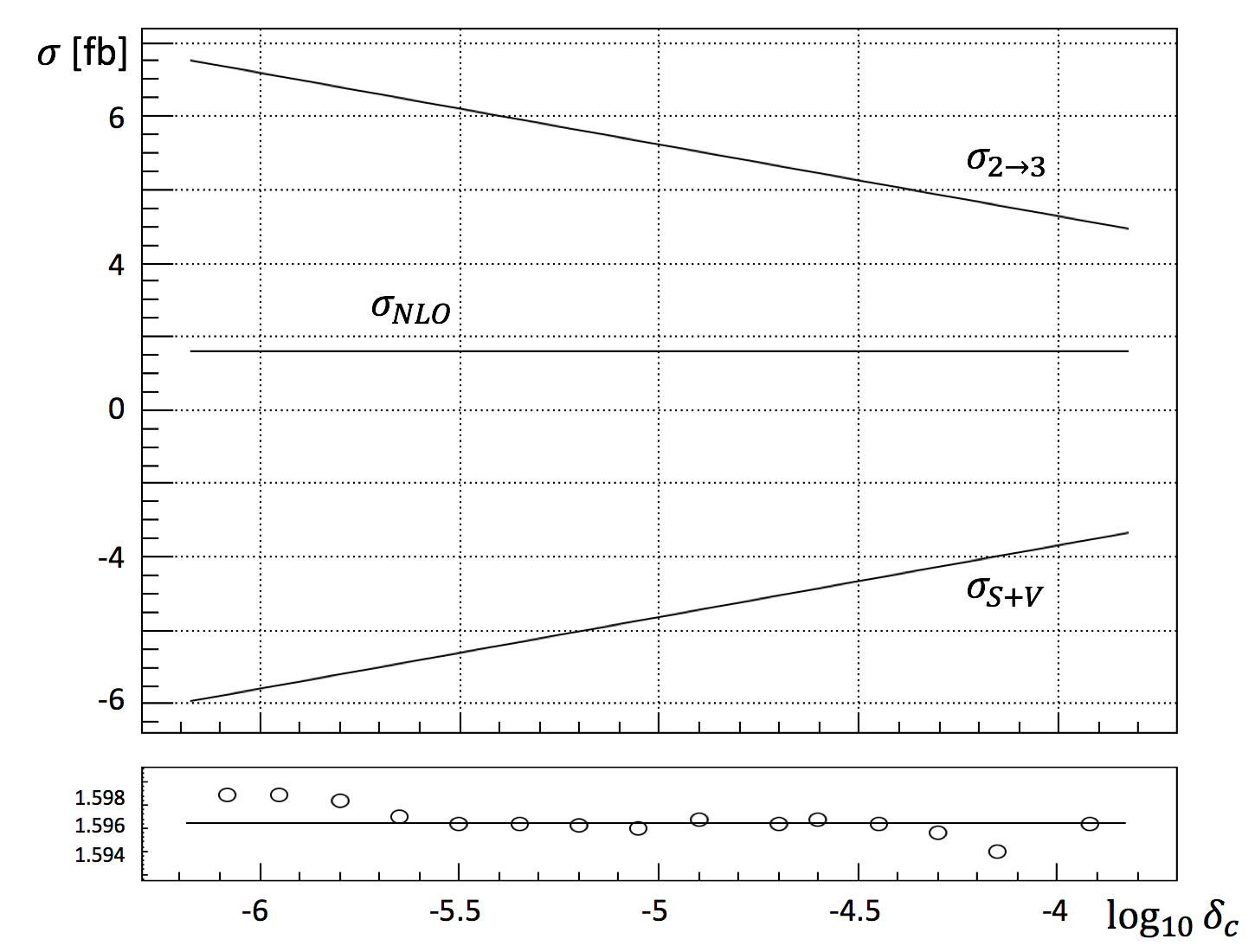}\\
\mycaption{Dependence of the NLO cross-section for the $pp\to CC$
on the soft cut-off $\delta_{\rm s}$ (left) and the collinear cut-off 
$\delta_{\rm c}$ (right). Both plots are
for a $pp$ center-of-mass energy of $\sqrt{s}=14\tev$, coloron mass $M=1\tev$,
and renormalization and factorization scales $\mu = \mu_{\rm F} = M$.
Furthermore, in the left (right) panel, the fixed value $\delta_{\rm c} =
10^{-5}$ ($\delta_{\rm s} = 10^{-3.5}$) has been used.
\label{fig:cutoff}}
\end{figure}

As a first consistency check, the
independence of the total NLO cross-section on the soft and collinear slicing
cut-offs, $\delta_{\rm s}$ and $\delta_{\rm c}$ is shown in
Fig.~\ref{fig:cutoff}. The figure depicts two separate plots for the dependence
on $\delta_{\rm s}$ and $\delta_{\rm c}$,
respectively. It can be seen that the combined virtual, soft and collinear
contributions ($\sigma_{\rm S+V}$) and the hard real emission contribution
($\sigma_{2\to 3}$) are separately logarithmically dependent on $\delta_{\rm s}$
and $\delta_{\rm c}$, but this dependence cancels in the sum $\sigma_{\rm NLO} =
\sigma_{\rm S+V}+\sigma_{2\to 3}$. The remaining power contributions,
proportional to $\delta_{\rm s}^n$ and $\delta_{\rm c}^n$, are negligibly small
for all practical purposes if the
cut-off parameters are smaller than about $10^{-3}$ and $10^{-4}$, respectively.

Note that the plots in Fig.~\ref{fig:cutoff} are subject to statistical errors
from the Monte-Carlo integration over initial parton momentum fractions and
final-state phase space. However, the
cancellation of soft and collinear logarithms happens already point-by-point for
the fully differential cross-section, after integration over only the
one-particle phase-space of the massless final-state parton in $\sigma_{2\to
3}$. Therefore, the accuracy of the cancellation of the $\delta_{\rm s}$
and $\delta_{\rm c}$ dependence is very high, as shown in the
lower boxes of the Fig.~\ref{fig:cutoff}.

The optimal choice of the cut-off parameters needs to strike a balance between
two constraints: \emph{(i)} The non-logarithmic
power contributions, proportional to $\delta^n_{\rm s,c}$, are minimized by
choosing each cut-off parameter as small as possible, whereas \emph{(ii)} the
statistical error for the $2 \to 3$ phase-space integration increases if
$\delta_{\rm s,c}$ are too small.
For the remainder of this section, we use $\delta_{\rm s} = 5 \times
10^{-4}$ and $\delta_{\rm c} = 5 \times 10^{-5}$.

\begin{figure}[t]
\centering
\includegraphics[width=10cm, trim=0 50 0 0, clip=true]{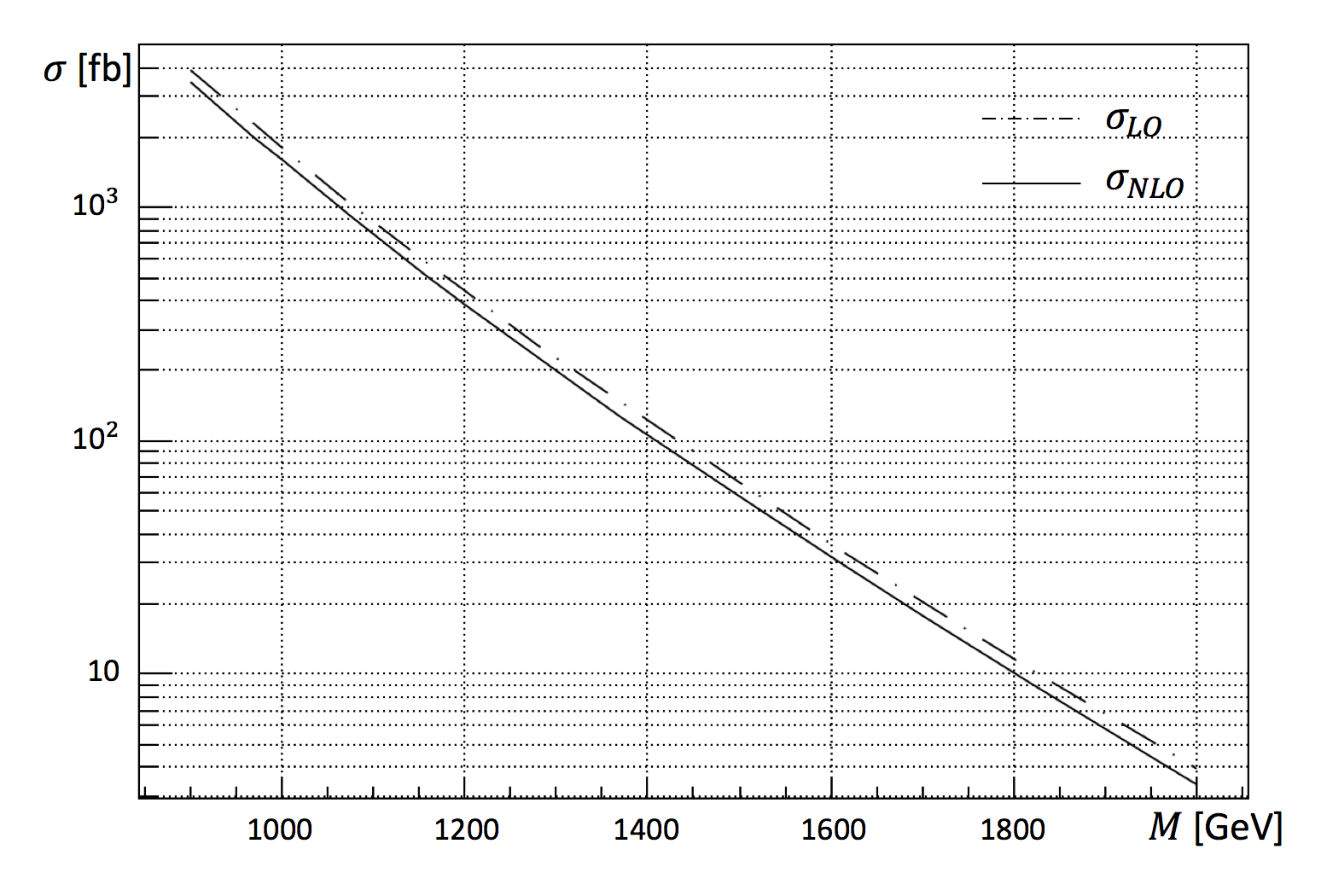}\\
\includegraphics[width=10.1cm, trim=-6 0 0 25, clip=true]{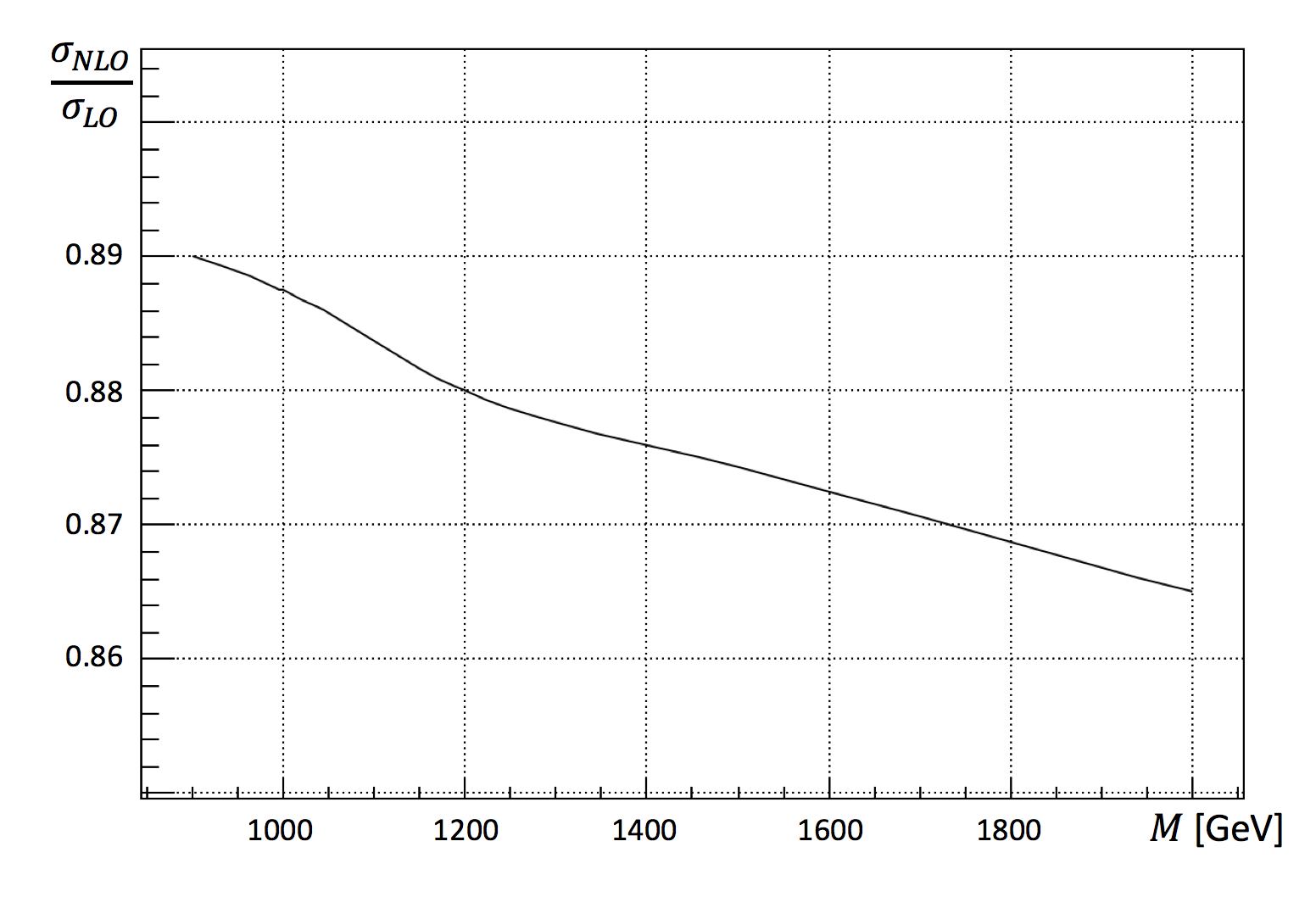}
\vspace{-1ex}
\mycaption{Total LO and NLO coloron pair production cross-sections as function of the
coloron mass $M$, for $\sqrt{s}=14\tev$ and $\mu = \mu_{\rm F} = M$. The mass splitting between the colorons
and $\cal P$-odd quarks has been set to the value predicted by mUED, see text
and eq.~\eqref{eq:delm}. The lower panel shows the ratio between NLO and LO
cross-sections.
\label{fig:xsecm}}
\end{figure}

\medskip\noindent
In Fig.~\ref{fig:xsecm}, the LO and NLO total cross-sections are shown as a
function of the coloron mass $M$. For this plot, the mass of the $\cal P$-odd
quarks has been fixed according to the mUED prediction, i.e.\ $M_\Psi = M -
\Delta M$, where $\Delta M$ is the mass splitting due to boundary terms in mUED
\cite{uedmass}
\begin{align}
\Delta M = M\,\frac{11\as}{16\pi} \, \ln \frac{\Lambda^2}{\mu^2}
\label{eq:delm}
\end{align}
Since $\Delta M$ is a one-loop contribution itself, we neglect it inside the
${\cal O}(\as)$ corrections to the cross-section and set $M_\Psi = M$ there. For the UV
cut-off of mUED we choose $\Lambda = 20 \, M$.

In the lower part of the figure, the $K$-factor $\sigma_{\rm NLO}/\sigma_{\rm LO}$ of
the NLO and Born cross-sections is shown. As evident from this plot, the
$K$-factor depends only mildly on $M$ and amounts to about 0.88.
It is interesting to note that the NLO contributions are negative in all three
subprocesses, $q\bar{q} \to CC +X$, $gg \to CC +X$, and $qg/\bar{q}g \to CC +
X$, the latter of which is only generated by $2\to 3$ real emission diagrams and
is turned negative due to the PDF renormalization.
While the overall correction is relatively modest, and of a typical magnitude for
high-energy QCD processes, it is nevertheless relevant for accurately evaluating current
limits and the discovery potential of the LHC for mUED and related models
\cite{uedlhc}.

\begin{figure}[tbp]
\centering
\vspace{-1em}
\includegraphics[width=12cm]{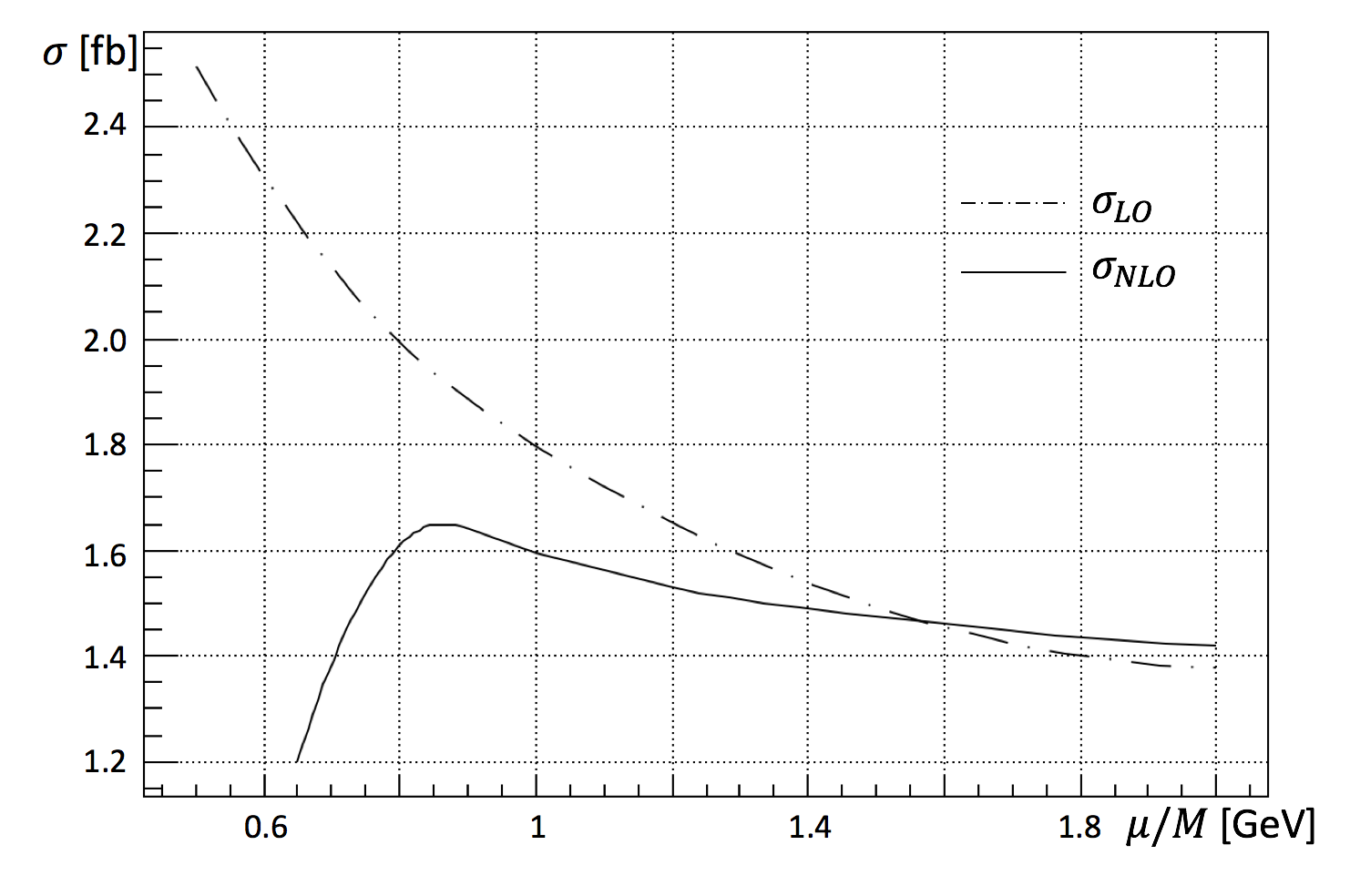}
\vspace{-1ex}
\mycaption{Dependence of the total LO and NLO coloron pair production cross-sections 
on the combined renormalization and factorization scale $\mu =
\mu_{\rm F}$. The plot is based on the $pp$ center-of-mass energy
$\sqrt{s}=14\tev$, mass $M=1\tev$, and mass splitting $M-M_\Psi$ as predicted by
mUED, see text and eq.~\eqref{eq:delm}.
\label{fig:scale}}
\end{figure}

In addition, the computation of the NLO QCD corrections serves to reduce the
theoretical uncertainty from the renormalization and factorization scale
dependence. This is demonstrated in Fig.~\ref{fig:scale}, where the two scales
have been varied in parallel, $\mu=\mu_{\rm F}$. Considering the range $0.75 <
\mu/M < 1.5$, the LO cross-section changes by about ${}^{+15\%}_{-17\%}$,
which is reduced to ${}^{+5\%}_{-8\%}$
for the NLO cross-section. 
Note that the dominant source of uncertainty stems from the
renormalization scale, whereas the factorization scale by itself has a
subdominant effect.

\begin{figure}[tbp]
\centering
\vspace{-1em}
\includegraphics[width=12cm]{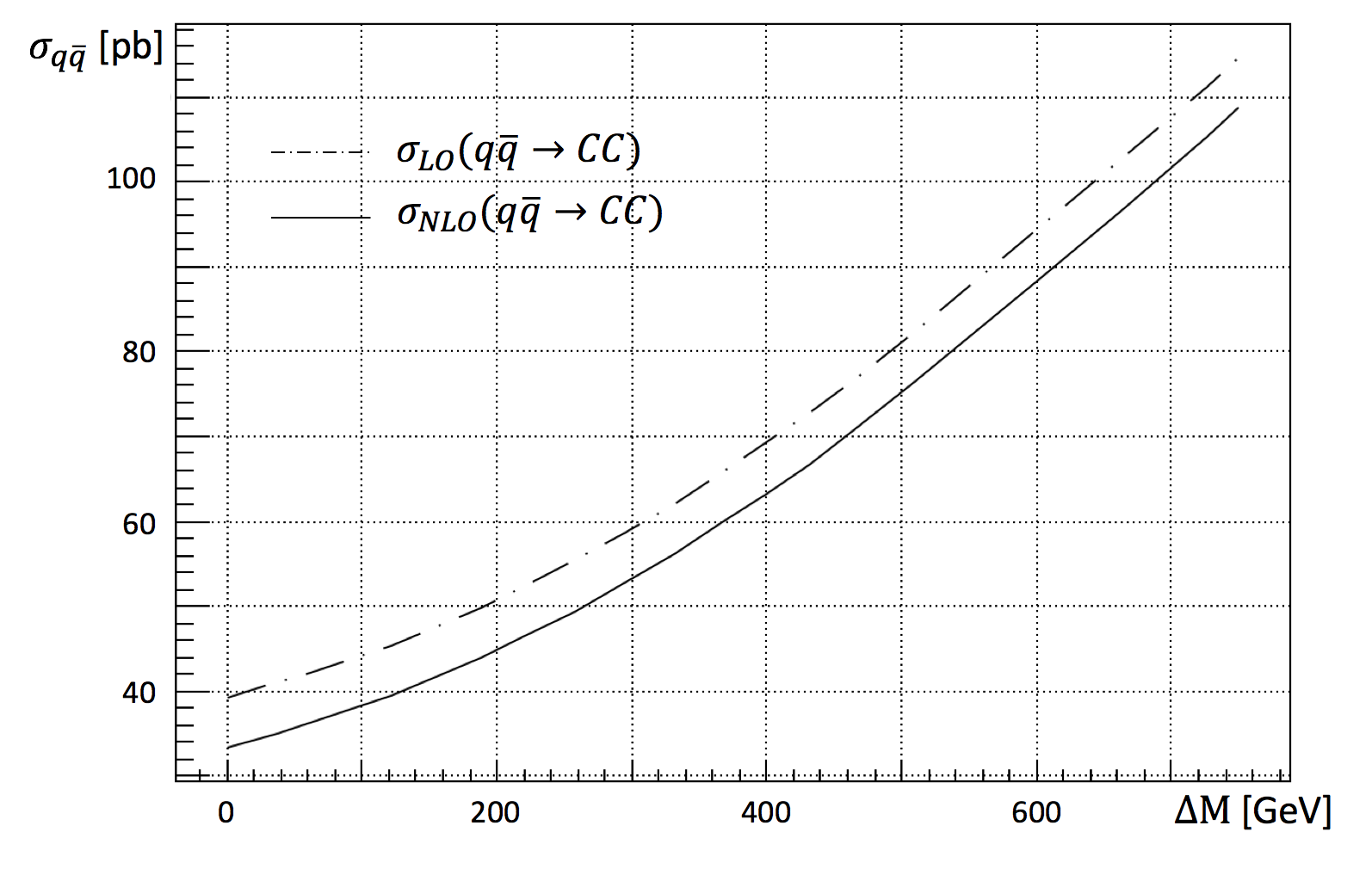}
\vspace{-1ex}
\mycaption{Total LO and NLO coloron pair production cross-sections 
as function of quark-coloron mass splitting $\Delta M = M - M_\Psi$. The other
input parameters have been set to $\sqrt{s}=14\tev$, mass $M=1\tev$, and
$\mu=\mu_{\rm F} =M$.
\label{fig:delm}}
\end{figure}

In Fig.~\ref{fig:delm}, we also show how the cross-section changes when the mass
splitting $\Delta M = M - M_\Psi$ is modified from the mUED prediction. Note
that the $gg$ channel does not depend on this parameter at tree-level, and we
neglect the mass splitting within the one-loop corrections. Therefore, only the
$q\bar{q}$ channel is shown in Fig.~\ref{fig:delm}. 
We restrict ourselves to the mass ordering $M_\Psi < M$, to avoid the situation
where the heavy quarks may become resonant in the subprocess $qg \to CCq$,
i.e.\ $qg \to C\Psi$ production with the subsequent decay $\Psi \to Cq$. This
would correspond to a different process than than the one studied in this paper
and is left for future work.
As evident from Fig.~\ref{fig:delm}, the $q\bar{q} \to CC$ subprocess depends very sensitively on  $\Delta M$.
However, since the $gg$ channel is dominant, the total cross-section varies
only by a few percent for reasonable values of the mass splitting.

\begin{figure}[tbp]
\centering
\vspace{-1em}
\includegraphics[width=12cm]{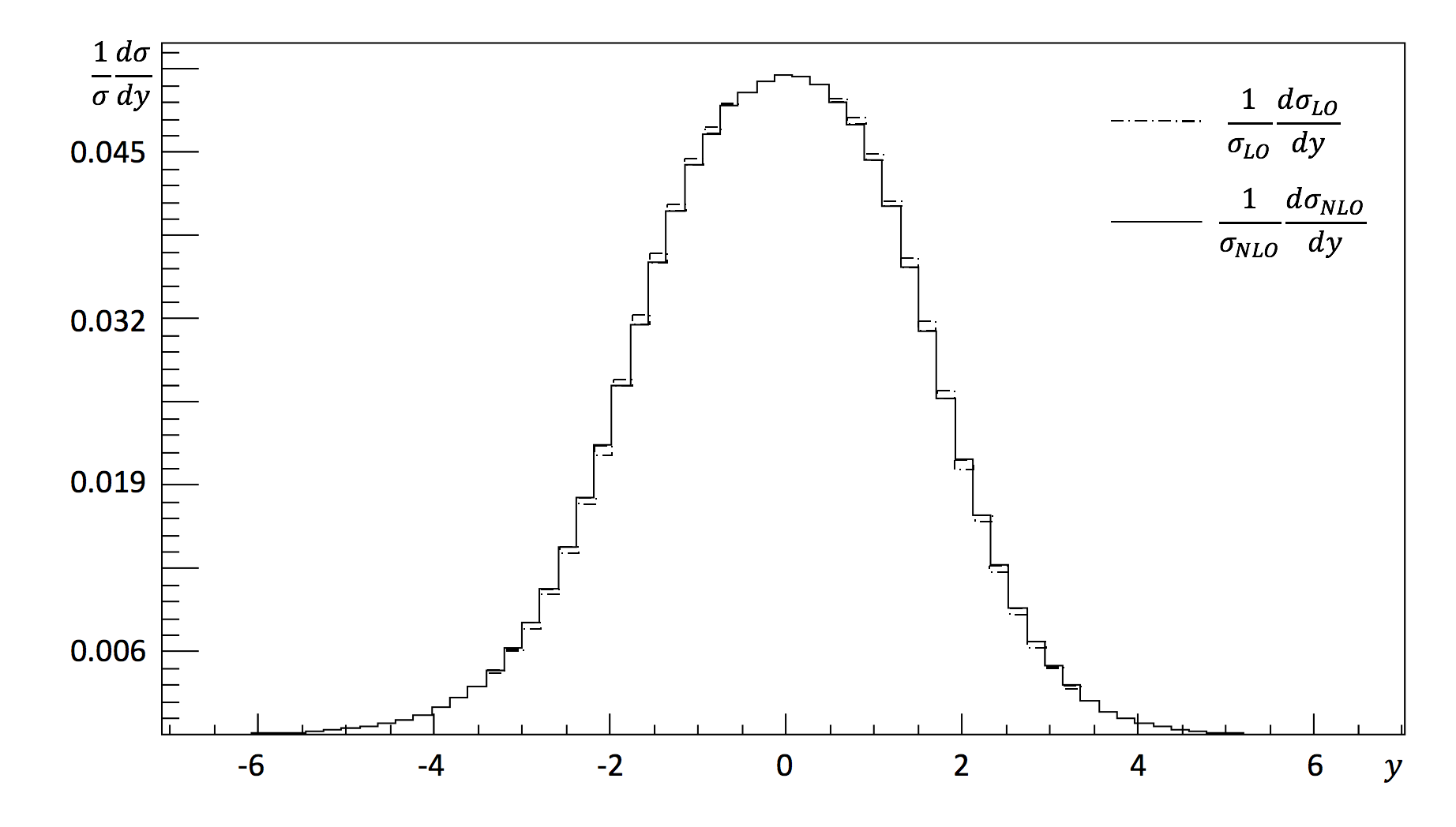}
\vspace{-1ex}
\mycaption{Differential cross-section for coloron pair production in terms of
rapidity at LO and NLO accuracy, for $\sqrt{s}=14\tev$, mass $M=1\tev$, and
$\mu=\mu_{\rm F} =M$.
\label{fig:diffy}}
\end{figure}

\medskip\noindent Finally, Fig.~\ref{fig:diffy} displays the impact of the NLO
corrections on the differential cross-section in terms of the rapidity $y \equiv
\frac{1}{2} \ln \frac{E+p_{\rm L}}{E-p_{\rm L}}$. Here $E$ and $p_{\rm L}$ are
the energy and longitudinal momentum of one of the final-state colorons. Since,
after summing over colors, we have two identical colorons in the final state,
the  rapidity distribution is symmetric. As one can see from the figure, the
effect of the NLO corrections results in a slight enhancement of the tails of
the rapidity distribution relative to the central region. This can be partially
understood from a simple kinematic effect, since the recoil against extra
radiated partons causes a broadening of the rapidity distribution.

\section{Conclusions}
\label{sc:concl}

\noindent The production of colored new physics particles at the LHC may be
subject to sizeable QCD corrections. In this article, results for the NLO
corrections to the pair production of color-octet vector bosons have been
presented. Such new vector bosons appear, for example, in coloron models or
models with extra space dimensions. There are characteristic versions of these
models where the single production of color-octet vector bosons is forbidden by
a parity symmetry, such as an exchange symmetry for coloron models and
Kaluza-Klein parity for extra dimensional models. For concreteness, this paper
focuses on a two-site coloron model, which is based on two copies of a
non-linear sigma model for the gauge sector. In addition, the presence of the
exchange symmetry requires the introduction of heavy partners to the SM quarks.
This model can serve as a
gauge-invariant low-energy effective description of the minimal universal extra
dimension (mUED) model.

The renormalization of the two-site coloron model involves several peculiarities
that do not occur for models with colored particles of spin less than one. For
instance, the couplings of the SM gluon and the massive coloron are identical at
tree-level, but they receive different counterterms at higher orders. In
addition, the broken gauge symmetry of the massive vector boson
requires the introduction of a counterterm for the symmetry-breaking vacuum
expectation value. This may be surprising at first glance, given that
the symmetry-breaking mechanism is not specified in the non-linear sigma model,
but in fact this counterterm can be uniquely determined from the Goldstone
self-energy.

The calculation of the NLO corrections presented in this paper is based on a
largely automated computer implementation, using
publicly available packages supplemented by in-house routines. For the
combination of virtual loop corrections and real radiation contributions, the
phase-space slicing method has been employed.
Several checks of the results have been performed.

It is found that for the standard choice of the renormalization scale, $\mu=M$,
where $M$ is the coloron mass, the NLO correction has a relatively modest impact
on the coloron pair production cross-section. The total NLO cross-section is
11--14\% smaller than the LO result for values of $M$ between 1 and 2~TeV. At
the same time, the dependence of the cross-section on the renormalization scale
is significantly reduced, by a factor of 2--3. By studying the rapidity
distribution it is furthermore observed that the NLO contribution cannot be
characterized by a simple global K-factor, but instead the K-factor is slightly
smaller in the central rapidity region and slightly larger for large absolute
values of rapidity.


\section*{Acknowledgments}

\noindent
The authors thank Z.~Qian for help with the Monte-Carlo integrator
routine from Ref.~\cite{Buckley:2014ana}.
This work has been supported in part by the National Science Foundation under
grant no.\ PHY-1519175.



\appendix

\section{Feynman rules of the two-site coloron model}
\label{sc:feynr}

This appendix lists the tree-level Feynman rules of the two-site symmetric
coloron model. The following notation is used:

\medskip
\begin{tabular}{ll}
$i,j,...$ & color indices in the fundamental representation \\
$A,B,...$ & color indices in the adjoint representation \\
$p_X$ & incoming momentum of the particle with color index $X$ \\
$\psi$ & generic SM quark \\
$\Psi$ & generic $\cal P$-odd quark \\
$Q$ & SU(2)-doublet $\cal P$-odd quark \\
$U$ & SU(2)-singlet $\cal P$-odd quark \\
$T,T'$ & $\cal P$-odd top partners, see eq.~\eqref{eq:tmix} \\
$\theta_T$ & mixing angle defined in eq.~\eqref{eq:tpars} \\
$P_{L,R} = \frac{1}{2}(1\pm \gamma_5)$ \\
$\eta^{\mu\nu}$ & metric tensor, $(\eta^{\mu\nu}) = \text{diag}(1,-1,-1,-1)$
\\[1ex]
\it Line styles:
\\[1ex]
single solid & SM quark \\
double solid & $\cal P$-odd quark \\
spring & gluon \\
spring--solid & coloron \\
dashed & $\cal P$-odd Goldstone scalar \\
dotted & ghost
\end{tabular}

\subsection{Feynman rules involving quarks except for the top quark}

\begin{minipage}{0.25\linewidth}
    \includegraphics[width=\linewidth]{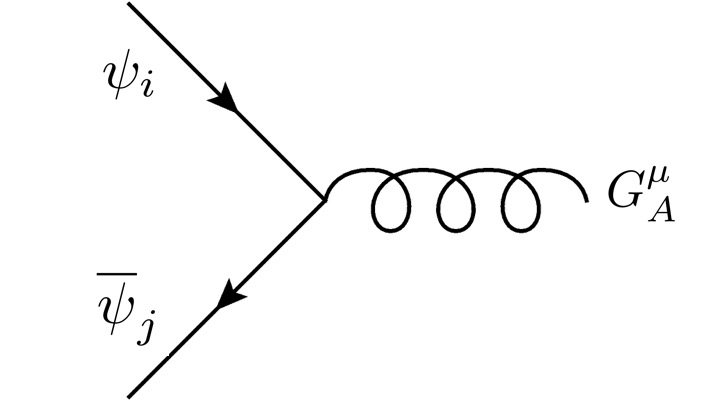}
\end{minipage}\hfill
\begin{minipage}{0.65\linewidth}
\begin{flalign}
-i\gs \gamma^\mu T^A_{ij}
&& \\ \notag \end{flalign}
\end{minipage}

\bigskip

\begin{minipage}{0.25\linewidth}
    \includegraphics[width=\linewidth]{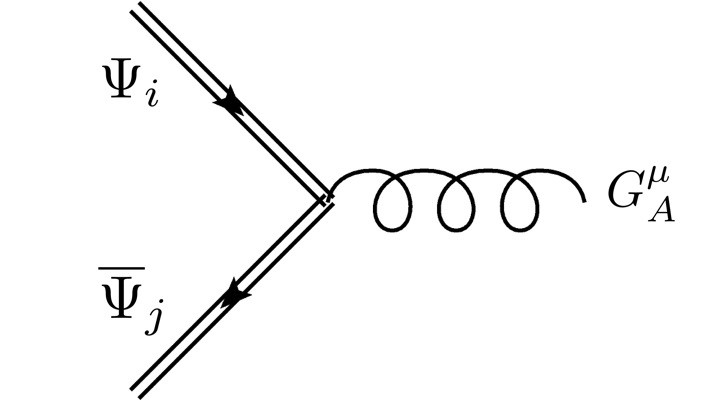}
\end{minipage}\hfill
\begin{minipage}{0.65\linewidth}
\begin{flalign}
-i\gs \gamma^\mu T^A_{ij}
&& \\ \notag \end{flalign}
\end{minipage}

\bigskip

\begin{minipage}{0.25\linewidth}
    \includegraphics[width=\linewidth]{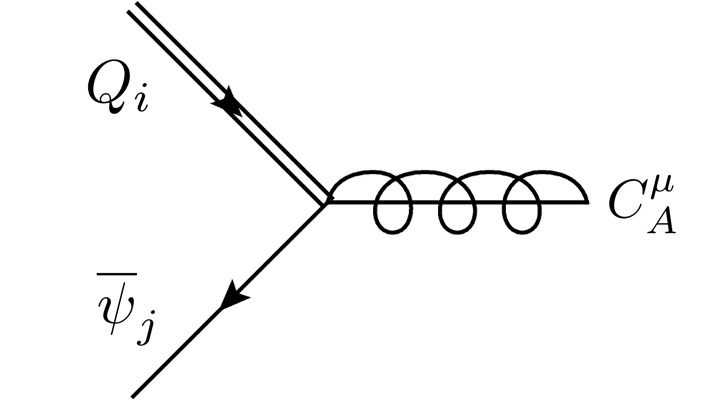}
\end{minipage}\hfill
\begin{minipage}{0.65\linewidth}
\begin{flalign}
-i\gs \gamma^\mu P_LT^A_{ij}
&& \\ \notag \end{flalign}
\end{minipage}

\bigskip

\begin{minipage}{0.25\linewidth}
    \includegraphics[width=\linewidth]{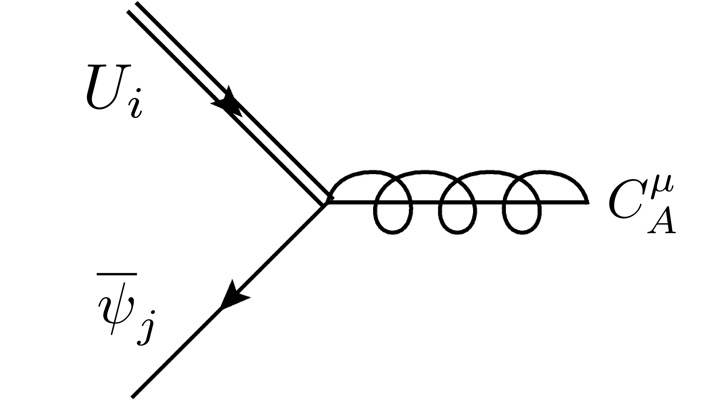}
\end{minipage}\hfill
\begin{minipage}{0.65\linewidth}
\begin{flalign}
i\gs \gamma^\mu P_RT^A_{ij} 
&& \\ \notag \end{flalign}
\end{minipage}

\bigskip

\begin{minipage}{0.25\linewidth}
    \includegraphics[width=\linewidth]{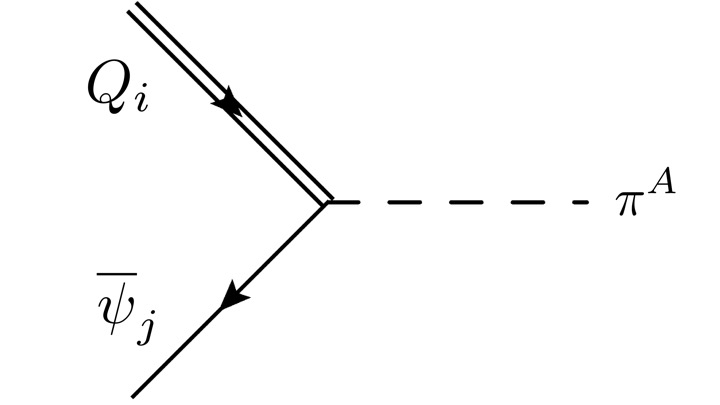}
\end{minipage}\hfill
\begin{minipage}{0.65\linewidth}
\begin{flalign}
-\gs P_RT^a_{ij}
&& \\ \notag \end{flalign}
\end{minipage}

\bigskip

\begin{minipage}{0.25\linewidth}
    \includegraphics[width=\linewidth]{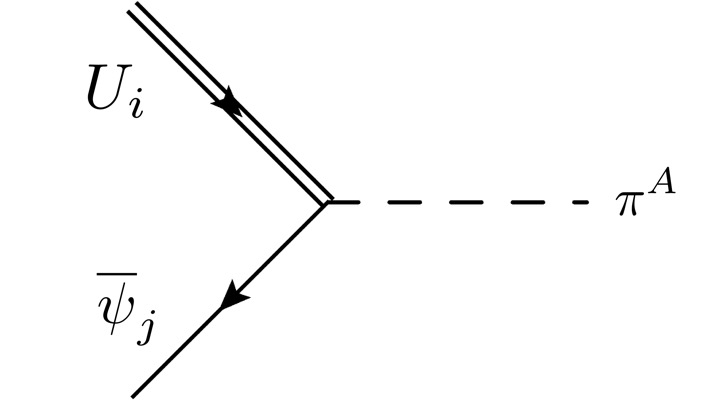}
\end{minipage}\hfill
\begin{minipage}{0.65\linewidth}
\begin{flalign}
\gs P_LT^a_{ij}
&& \\ \notag \end{flalign}
\end{minipage}

\subsection{Vertices involving the top quark}


\begin{minipage}{0.25\linewidth}
    \includegraphics[width=\linewidth]{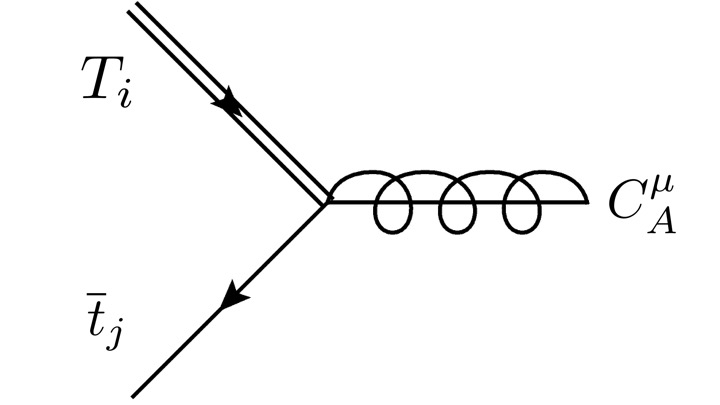}
\end{minipage}\hfill
\begin{minipage}{0.65\linewidth}
\begin{flalign}
-i\gs \gamma^\mu\left[\sin{\theta_T}P_R+\cos{\theta_T}P_L\right]T^a_{ij}
&& \\ \notag \end{flalign}
\end{minipage}

\bigskip

\begin{minipage}{0.25\linewidth}
    \includegraphics[width=\linewidth]{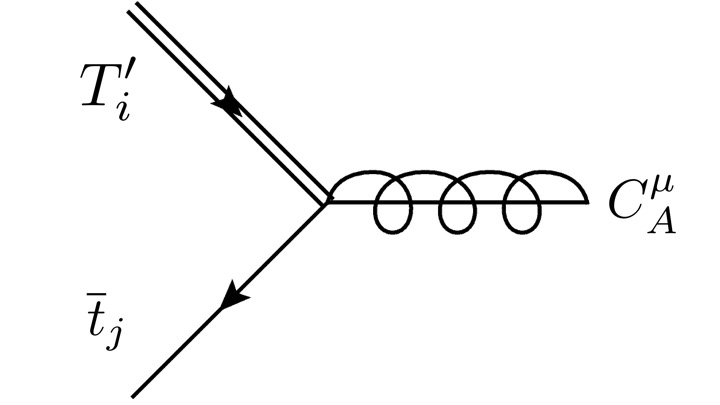}
\end{minipage}\hfill
\begin{minipage}{0.65\linewidth}
\begin{flalign}
i\gs \gamma^\mu\left[\sin{\theta_T}P_L+\cos{\theta_T}P_R\right]T^a_{ij}
&& \\ \notag \end{flalign}
\end{minipage}

\bigskip

\begin{minipage}{0.25\linewidth}
    \includegraphics[width=\linewidth]{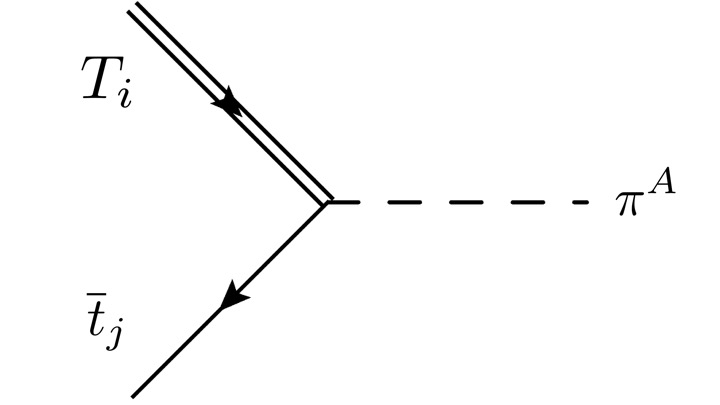}
\end{minipage}\hfill
\begin{minipage}{0.65\linewidth}
\begin{flalign}
\gs \left[\sin{\theta_T}P_L-\cos{\theta_T}P_R\right]T^a_{ij}
&& \\ \notag \end{flalign}
\end{minipage}

\bigskip

\begin{minipage}{0.25\linewidth}
    \includegraphics[width=\linewidth]{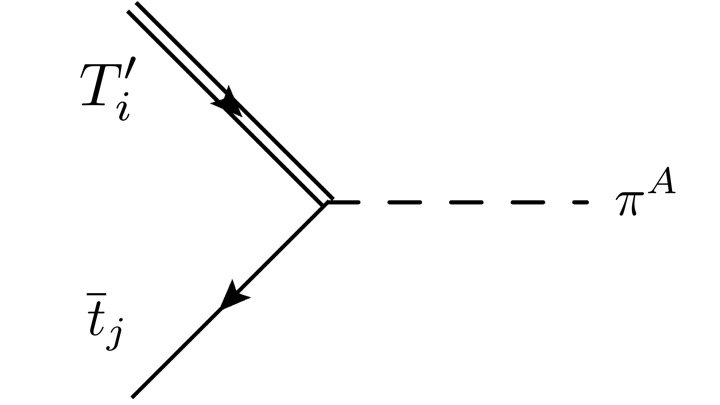}
\end{minipage}\hfill
\begin{minipage}{0.65\linewidth}
\begin{flalign}
-\gs \left[\sin{\theta_T}P_R-\cos{\theta_T}P_L\right]T^a_{ij}
&& \\ \notag \end{flalign}
\end{minipage}

\bigskip

\subsection{Three-point boson vertices}

\begin{minipage}{0.25\linewidth}
    \includegraphics[width=\linewidth]{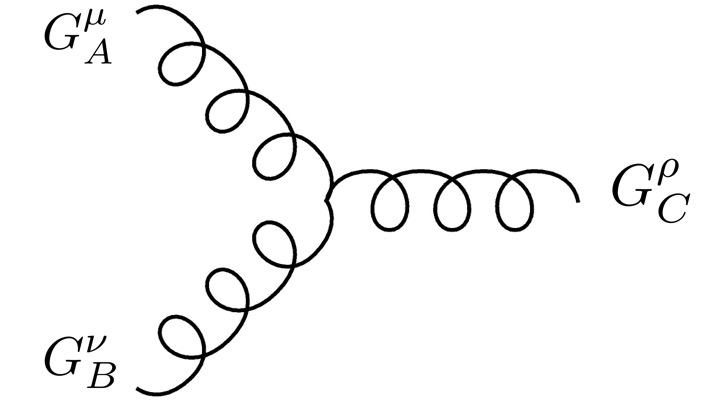}
\end{minipage}\hfill
\begin{minipage}{0.65\linewidth}
\begin{flalign}
\gs\left[\left(p_B-p_A\right)^{\rho}\eta^{\mu\nu}
        +\left(p_A-p_C\right)^{\nu}\eta^{\mu\rho}
	+\left(p_C-p_B\right)^{\mu}\eta^{\nu\rho}\right]f^{ABC}
&& \notag\\  \end{flalign}
\end{minipage}

\bigskip

\begin{minipage}{0.25\linewidth}
    \includegraphics[width=\linewidth]{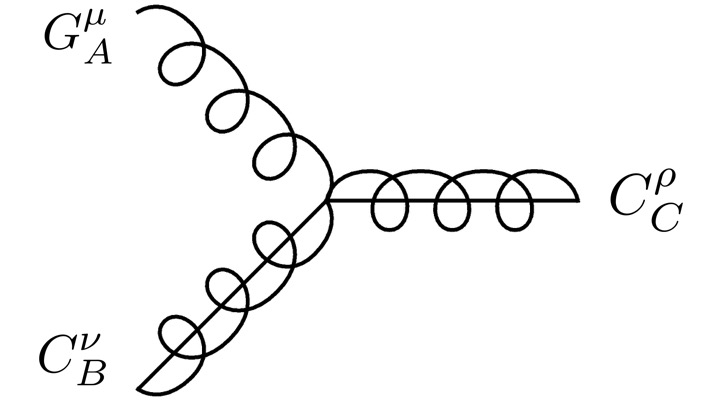}
\end{minipage}\hfill
\begin{minipage}{0.65\linewidth}
\begin{flalign}
\gs\left[\left(p_B-p_A\right)^{\rho}\eta^{\mu\nu}
        +\left(p_A-p_C\right)^{\nu}\eta^{\mu\rho}
	+\left(p_C-p_B\right)^{\mu}\eta^{\nu\rho}\right]f^{ABC}
&& \notag\\  \end{flalign}
\end{minipage}

\bigskip

\begin{minipage}{0.25\linewidth}
    \includegraphics[width=\linewidth]{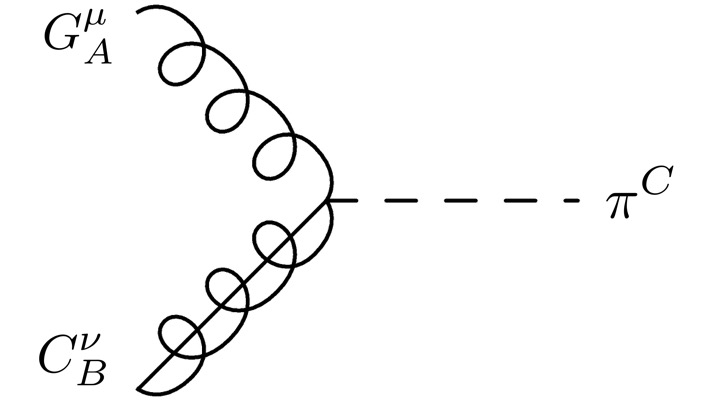}
\end{minipage}\hfill
\begin{minipage}{0.65\linewidth}
\begin{flalign}
-i\gs M\eta^{\mu\nu}f^{ABC}
&& \\ \notag \end{flalign}
\end{minipage}

\bigskip

\begin{minipage}{0.25\linewidth}
    \includegraphics[width=\linewidth]{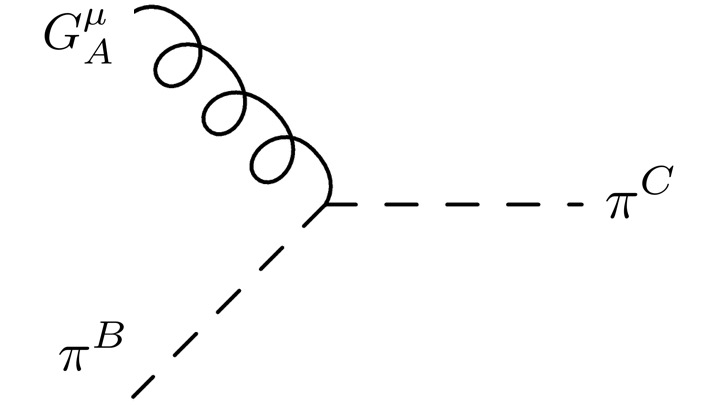}
\end{minipage}\hfill
\begin{minipage}{0.65\linewidth}
\begin{flalign}
\gs \left(p_B-p_C\right)^\mu f^{ABC}
&& \\ \notag \end{flalign}
\end{minipage}

\bigskip

\subsection{Feynman rules involving ghosts}

\begin{minipage}{0.25\linewidth}
    \includegraphics[width=\linewidth]{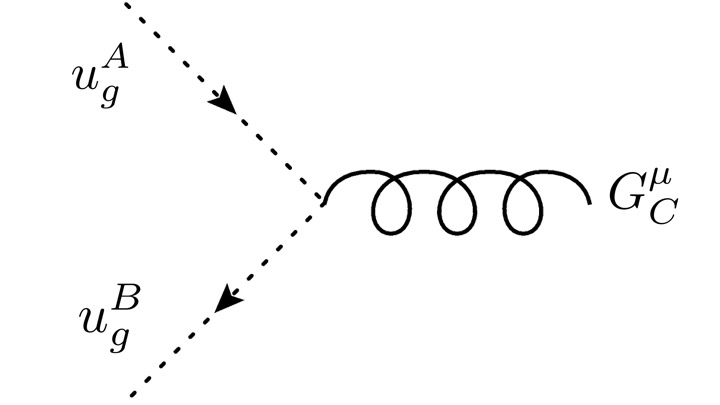}
\end{minipage}\hfill
\begin{minipage}{0.65\linewidth}
\begin{flalign}
-\gs p_B^\mu f^{ABC}
&& \\ \notag \end{flalign}
\end{minipage}

\bigskip

\begin{minipage}{0.25\linewidth}
    \includegraphics[width=\linewidth]{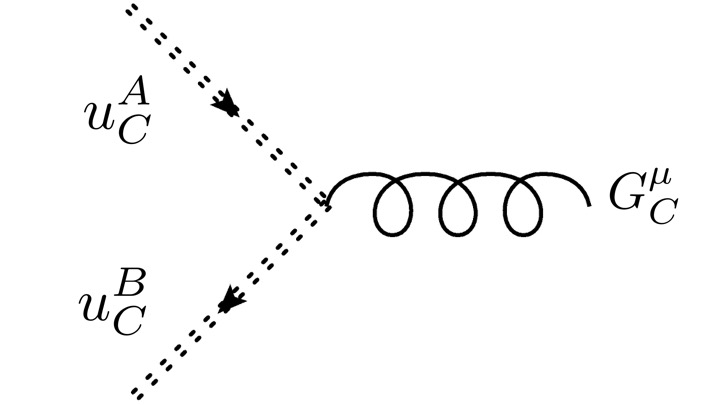}
\end{minipage}\hfill
\begin{minipage}{0.65\linewidth}
\begin{flalign}
-\gs p_B^\mu f^{ABC}
&& \\ \notag \end{flalign}
\end{minipage}

\bigskip

\begin{minipage}{0.25\linewidth}
    \includegraphics[width=\linewidth]{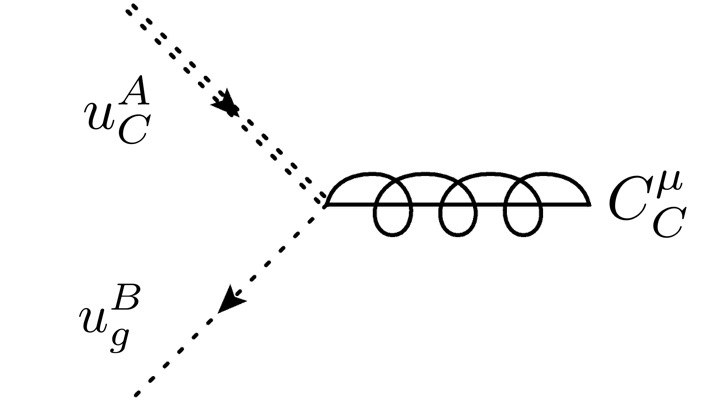}
\end{minipage}\hfill
\begin{minipage}{0.65\linewidth}
\begin{flalign}
-\gs p_B^\mu f^{ABC}
&& \\ \notag \end{flalign}
\end{minipage}

\bigskip

\begin{minipage}{0.25\linewidth}
    \includegraphics[width=\linewidth]{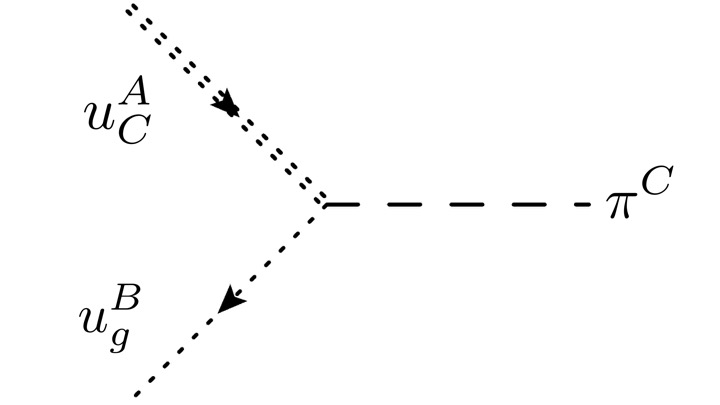}
\end{minipage}\hfill
\begin{minipage}{0.65\linewidth}
\begin{flalign}
-i\gs Mf^{ABC}
&& \\ \notag \end{flalign}
\end{minipage}

\bigskip

\subsection{Four-point boson vertices}

\begin{minipage}{0.25\linewidth}
    \includegraphics[width=\linewidth]{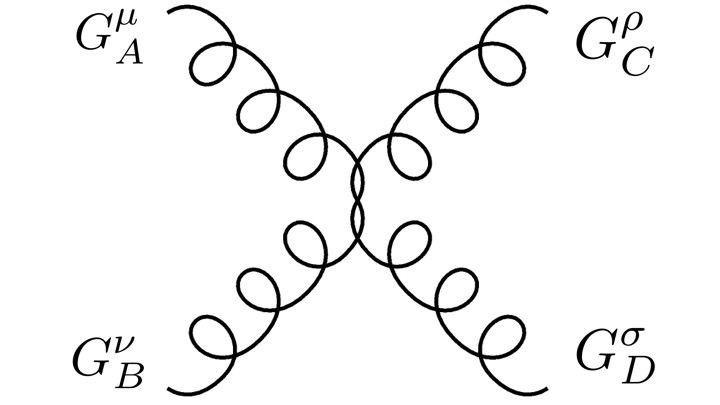}
\end{minipage}\hfill
\begin{minipage}{0.65\linewidth}
\begin{flalign}
-i\gs^2\bigl[&\eta^{\mu\nu}\eta^{\rho\sigma}\bigl(f^{ACE}f^{BDE}-f^{ADE}f^{CBE}\bigr)
\nonumber &\\
+&\eta^{\mu\rho}\eta^{\nu\sigma}\bigl(f^{ADE}f^{CBE}-f^{ABE}f^{DCE}\bigr)
\nonumber & \\
+&\eta^{\mu\sigma}\eta^{\nu\rho}\bigl(f^{ABE}f^{DCE}-f^{ACE}f^{BDE}\bigr)\bigr] &
\end{flalign}
\end{minipage}

\bigskip

\begin{minipage}{0.25\linewidth}
    \includegraphics[width=\linewidth]{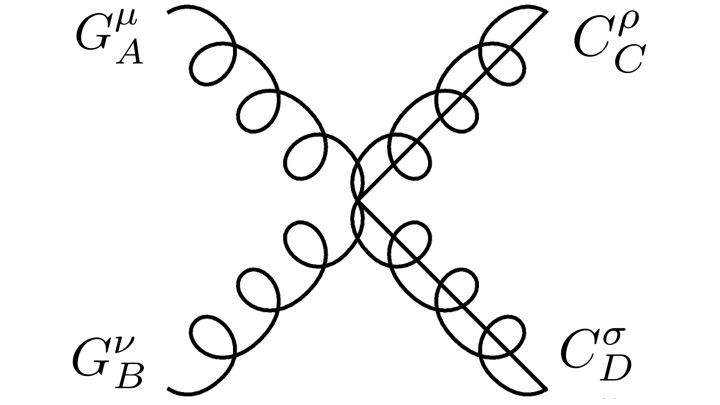}
\end{minipage}\hfill
\begin{minipage}{0.65\linewidth}
\begin{flalign}
-i\gs^2\bigl[&\eta^{\mu\nu}\eta^{\rho\sigma}\bigl(f^{ACE}f^{BDE}-f^{ADE}f^{CBE}\bigr)
\nonumber &\\
+&\eta^{\mu\rho}\eta^{\nu\sigma}\bigl(f^{ADE}f^{CBE}-f^{ABE}f^{DCE}\bigr)
\nonumber & \\
+&\eta^{\mu\sigma}\eta^{\nu\rho}\bigl(f^{ABE}f^{DCE}-f^{ACE}f^{BDE}\bigr)\bigr] &
\end{flalign}
\end{minipage}

\bigskip

\begin{minipage}{0.25\linewidth}
    \includegraphics[width=\linewidth]{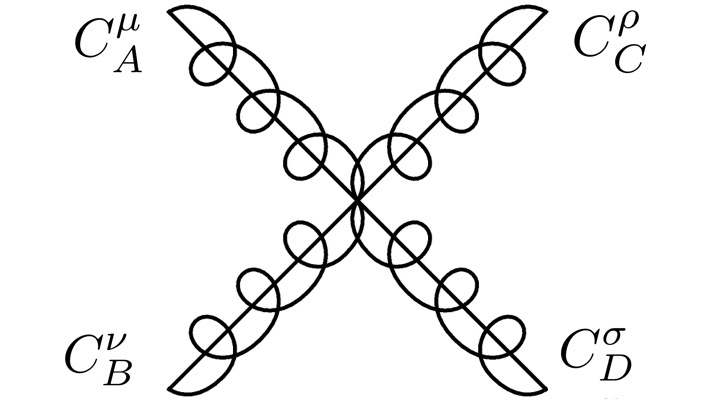}
\end{minipage}\hfill
\begin{minipage}{0.65\linewidth}
\begin{flalign}
-i\gs^2\bigl[&\eta^{\mu\nu}\eta^{\rho\sigma}\bigl(f^{ACE}f^{BDE}-f^{ADE}f^{CBE}\bigr)
\nonumber &\\
+&\eta^{\mu\rho}\eta^{\nu\sigma}\bigl(f^{ADE}f^{CBE}-f^{ABE}f^{DCE}\bigr)
\nonumber & \\
+&\eta^{\mu\sigma}\eta^{\nu\rho}\bigl(f^{ABE}f^{DCE}-f^{ACE}f^{BDE}\bigr)\bigr] &
\label{eq:cccc}
\end{flalign}
\end{minipage}

\bigskip

\begin{minipage}{0.25\linewidth}
    \includegraphics[width=\linewidth]{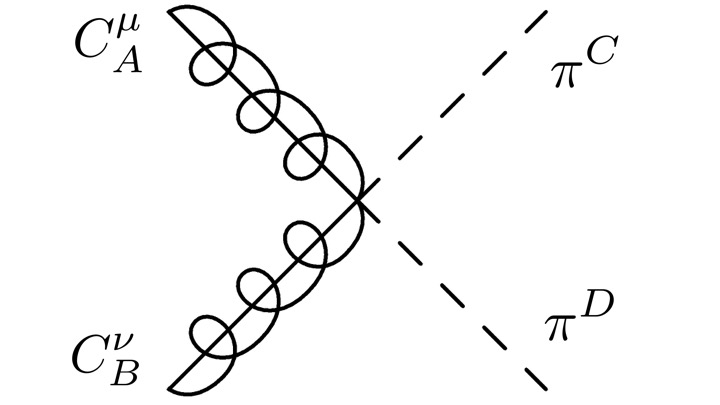}
\end{minipage}\hfill
\begin{minipage}{0.65\linewidth}
\begin{flalign}
i\gs ^2\eta^{\mu\nu}\left(f^{ACE}f^{BDE}+f^{BCE}f^{ADE}\right)
&& \\ \notag \end{flalign}
\end{minipage}

\bigskip

\begin{minipage}{0.25\linewidth}
    \includegraphics[width=\linewidth]{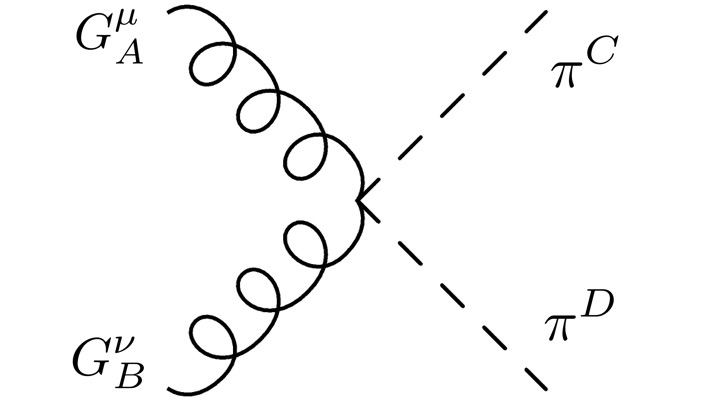}
\end{minipage}\hfill
\begin{minipage}{0.65\linewidth}
\begin{flalign}
i\gs ^2\eta^{\mu\nu}\left(f^{ACE}f^{BDE}+f^{BCE}f^{ADE}\right)
&& \\ \notag \end{flalign}
\end{minipage}

\bigskip

\noindent
Note that the Feynman rules in this appendix agree with those for KK-level--1
gluons and quarks in mUED, with the exception of \eqref{eq:cccc}, which
has an additional factor $\frac{3}{2}$ in mUED.


\end{document}